\documentclass[12pt]{amsart}

\copyrightinfo{2000}{American Mathematical Society}

\newcommand{\cA}{{\mathcal A}}
\newcommand{\cB}{{\mathcal B}}
\newcommand{\cD}{{\mathcal D}}
\newcommand{\cE}{{\mathcal E}}
\newcommand{\cL}{{\mathcal L}}

\newcommand{\bN}{{\mathbb N}}
\newcommand{\bZ}{{\mathbb Z}}
\newcommand{\bQ}{{\mathbb Q}}
\newcommand{\bR}{{\mathbb R}}
\newcommand{\bC}{{\mathbb C}}

\numberwithin{equation}{section}

\newtheorem{Theorem}{Theorem}[section]
\newtheorem{Lemma}{Lemma}[section]
\newtheorem{Corollary}{Corollary}[section]
\newtheorem{Definition}{Definition}[section]
\newtheorem{Remark}{Remark}[section]

\newtheorem{Proposition}{Proposition}[section]

\author{V.~M.~Shelkovich}
\address{Department of Mathematics, St.-Petersburg State Architecture
and Civil Engineering University, \ 2 Krasnoarmeiskaya 4, 190005,
St. Petersburg, \ Russia. \ Phone: +7\,(812)\,2517549 \,
Fax: +7\,(812)\,3165872}
\email{shelkv@vs1567.spb.edu}

\author{M.~Skopina}
\address{Department of Applied Mathematics and Control Processes,
St. Petersburg State University, \ Universitetskii pr.-35,
Petrodvorets, 198504 St. Petersburg, Russia. \ Phone: +7\,(812)\,51326090 \,
}
\email{skopina@MS1167.spb.edu}

\title[$p$-Adic Haar multiresolution analysis]
{$p$-Adic Haar multiresolution analysis and pseudo-differential
operators}

\thanks{The first author (V.~S.) is supported by
DFG Project 436 RUS 113/809 and Grant 05-01-04002-NNIOa of
Russian Foundation for Basic Research, the second author (M.~S.)
is supported by Grant 06-01-00457 of Russian Foundation
for Basic Research.}



\date{ }

\begin{document}

\begin{abstract}
The notion of {\em  $p$-adic multiresolution
analysis (MRA)} is introduced. We discuss a ``natural'' refinement
equation whose solution (a refinable function) is the
characteristic function of the unit disc. This equation reflects
the fact that the characteristic function of the unit disc is a
sum of $p$ characteristic functions of mutually disjoint discs of radius
$p^{-1}$. This refinement equation generates a MRA. The case $p=2$
is studied in detail. Our MRA is a $2$-adic analog of the real
Haar MRA. But in contrast to the real setting, the refinable
function generating our Haar MRA is $1$-periodic, which never
holds for real refinable functions. This fact implies that there
exist infinity many different $2$-adic orthonormal wavelet bases
in ${\cL}^2(\bQ_2)$ generated by the same Haar MRA. All of these
bases are described. We also constructed multidimensional $2$-adic
Haar  orthonormal bases for ${\cL}^2(\bQ_2^n)$ by means of the tensor
product of one-dimensional MRAs. A criterion for a multidimensional
$p$-adic wavelet to be an eigenfunction for a pseudo-differential
operator is derived. We proved also that these wavelets are
eigenfunctions of the Taibleson multidimensional fractional
operator. These facts create the necessary prerequisites
for intensive using  our bases in applications.
\end{abstract}

\maketitle

\section{Introduction}
\label{s1}

According to the well-known Ostrovsky theorem, {\it any nontrivial
valuation on the field ${\bQ}$ is equivalent either to the real
valuation $|\cdot|$ or to one of the $p$-adic valuations $|\cdot|_p$}.
This $p$-adic norm $|\cdot|_p$ is defined as follows:
if an arbitrary rational number $x\ne 0$ is represented as
$x=p^{\gamma}\frac{m}{n}$, where $\gamma=\gamma(x)\in \bZ$ and
the integers $m$, $n$ are not divisible by $p$, then
\begin{equation}
\label{1}
|x|_p=p^{-\gamma}, \quad x\ne 0, \qquad |0|_p=0.
\end{equation}
The norm $|\cdot|_p$ satisfies the strong triangle inequality
$|x+y|_p\le \max(|x|_p,|y|_p)$ and is non-Archimedean.
The field $\bQ_p$ of $p$-adic numbers is defined as the
completion of the field of rational numbers $\bQ$ with respect
to the norm $|\cdot|_p$.

Thus there are two equal in rights universes: the real universe
and the $p$-adic one. The latter has a specific and unusual
properties. Nevertheless, there are a lot of papers where different
applications of $p$-adic analysis to physical problems, stochastics,
cognitive sciences and psychology are studied~\cite{Ar-Dr-V}--
~\cite{Bik-V},~\cite{Kh1}--~\cite{Koch3},~\cite{Vl-V-Z}--~\cite{V2}
(see also the references therein). In view of the Ostrovsky theorem,
such investigations are not only of great interest in itself, but
lead to applications and better understanding of similar problems
in {\it usual} mathematical physics.

Since there exists a $p$-adic analysis connected with the mapping
$\bQ_p$ into $\bQ_p$ and an analysis connected with the mapping
$\bQ_p$ into the field of complex numbers $\bC$, one considers
two corresponding types of $p$-adic physical models.
For the $p$-adic analysis related to the mapping $\bQ_p \to \bC$,
the operation of differentiation is {\it not defined\/},
and as a result, large number of models connected with $p$-adic
differential equations use pseudo-differential operators (see
the above-mentioned papers and books). In particular, fractional
operator $D^{\alpha}$ are extensively used in applications.
A very important fact that the eigenfunctions of a one-dimensional
fractional operator $D^{\alpha}$ form an orthonormal basis for
${\cL}^2(\bQ_p)$ was observed by V.S.~Vladimirov, I.V.~Volovich,
E.I.~Zelenov (see~\cite{Vl-V-Z}).
S.~V.~Kozyrev~\cite{Koz0} found an orthonormal
compactly supported $p$-adic wavelet basis for ${\cL}^2(\bQ_p)$:
\begin{equation}
\label{62.0-1}
\theta_{k;j a}(x)=p^{-\gamma/2}\chi_p\big(p^{-1}k(p^{j}x-a)\big)
\Omega\big(|p^{j}x-a|_p\big), \quad x\in \bQ_p,
\end{equation}
$k=1,2,\dots,p-1$, $j\in \bZ$, $a\in I_p=\bQ_p/\bZ_p$.
Wavelets (\ref{62.0-1}) are also eigenfunctions of the one-dimensional
fractional operator $D^{\alpha}$:
$$
D^{\alpha}\theta_{k;j a}(x)=p^{\alpha(1-j)}\theta_{k;j a}(x),
\quad x\in \bQ_p, \qquad \alpha\in \bC.
$$
Some wavelet-type systems generalizing (\ref{62.0-1}) were suggested
by S.~V.~Kozyrev \cite{Koz1},~\cite{Koz2}, A.~Yu.~Khrennikov
and S.~V.~Kozyrev~\cite{Kh-Koz1}, \cite{Kh-Koz2}, J.~J.~Benedetto and
R.~L.~Benedetto~\cite{Ben-Ben},  R.~L.~Benedetto~\cite{Ben1}.
Multidimensional $p$-adic bases obtained by direct multiplying out
the Kozyrev's wavelets (\ref{62.0-1}) were considered in~\cite{Al-Kh-Sh3}.
The authors of~\cite{Kh-Sh1} found the following new type of $p$-adic
wavelet basis:
\begin{equation}
\label{62.0-1*}
\theta_{s;j a}^{(m)}(x)=p^{-j/2}\chi_p\big(s(p^{j}x-a)\big)
\Omega\big(|p^{j}x-a|_p\big), \quad x\in \bQ_p,
\end{equation}
where $m\ge 1$ is a {\it fixed} positive
integer;
$s=p^{-m}\big(s_{0}+s_{1}p+\cdots+s_{m-1}p^{m-1}\big)$,
$s_r=0,1,\dots,p-1$, \ $r=0,1,\dots,m-1$, \ $s_0\ne 0$;
\ $j\in \bZ$, $a\in I_p$.
It turned out that wavelets (\ref{62.0-1}) and their generalizations
are eigenfunctions of $p$-adic pseudo-differential
operators~\cite{Al-Kh-Sh3}--~\cite{Al-Kh-Sh5},~\cite{Kh-Koz1},
~\cite{Kh-Koz2},~\cite{Kh-Sh1},~\cite{Koz0} --~\cite{Koz2}.
Moreover,  a necessary and sufficient conditions for a class of
$p$-adic pseudo-differential operators (\ref{64.3}) (including
fractional operator (\ref{59**})) to have wavelets (\ref{62.0-1})
and (\ref{62.0-1*}) as eigenfunctions was
derived in~\cite{Al-Kh-Sh3},~\cite{Kh-Sh1}.
So,  wavelets play an important role for application of $p$-adic
analysis and gives a new powerful technique for solving $p$-adic
problems. Nevertheless, in the  cited papers, there was no any attempt
to create a theory describing common properties of wavelet bases and
giving methods for their finding. The goal of this paper is to start
development of such a theory.

It's interesting to compare appearing  first wavelets in $p$-adic
analysis with the history of the wavelet theory in real analysis.
In 1910 Haar~\cite{25} constructed an orthogonal basis for
$\cL^2(\bR)$ consisting of the dyadic shifts and scales of one
piecewise constant function. A lot of mathematicians actively
studied Haar basis, different kinds of generalizations were
introduced, but during almost the whole century nobody could find
another wavelet function (a function whose shifts and scales form
an orthogonal basis). Only in early nineties a general scheme  for
construction of wavelet functions was developed. This scheme is based
on the notion of multiresolution analysis (MRA in the sequel)
introduced by Y.~Meyer and S.~Mallat~\cite{13}, \cite{15},
\cite{18}. Smooth compactly supported wavelet functions were found
in this way, which has been very important for various engineering
applications. In the present paper we introduce MRA in $\cL^2(\bQ_p)$ and
study a concrete MRA for $p=2$ being an analog of Haar MRA in
$\cL^2(\bR)$. The same scheme as  in the real setting leads to a
Haar basis. It turned out that this Haar basis coincides with
Kozyrev's wavelet system (\ref{62.0-1}). However, 2-adic Haar MRA is not an
identical copy of its real analog.  We proved that, in contrast to Haar MRA in
$\cL^2({\bR})$, there exist {\it infinity many
different Haar orthogonal bases} for $\cL^2({\bQ}_2)$ generated by
the same MRA.

The paper is organized as follows.

In Sec.~\ref{s2}, we recall some facts from the $p$-adics.
The basic results on the theory of the Bruhat--Schwartz distributions
are given in Subsec.~\ref{s2.1} (see~\cite{G-Gr-P},~\cite{Taib1},
\cite{Taib3},~\cite{Vl-V-Z}). Some facts from the theory
of the $p$-adic space $\Phi'(\bQ_p^n)$ of Lizorkin distributions~\cite{Al-Kh-Sh3}
are summarized in Subsec.~\ref{s2.2}.
In Subsec.~\ref{s7.1},~\ref{s7.2}, we recall some facts~\cite{Al-Kh-Sh3}
related to the multidimensional pseudo-differential operators (\ref{64.3})
in the space of Lizorkin distributions $\Phi'(\bQ_p^n)$. In particular,
the multidimensional fractional operators introduced by Taibleson
~\cite[\S2]{Taib1},~\cite[III.4.]{Taib3} are
discussed. The spaces $\Phi'(\bQ_p^n)$ is a ``natural''
domain for the class of pseudo-differential operators
(\ref{64.3}). The spaces $\Phi'(\bQ_p^n)$  are {\it invariant\/}
under our pseudo-differential operators. It is appropriate to
mention here that the class of our operators includes the
pseudo-differential operators studied in~\cite{Koch3},~\cite{Z1},~\cite{Z2}.

In Sec.~\ref{s4},   a notion of $p$-adic  MRA is introduced (Definition~\ref{de1}).
In Subsec.~\ref{s4.2}, we discuss  {\it refinement equation}
(\ref{62.0-3}):
$$
\phi(x)=\sum_{r=0}^{p-1}\phi\Big(\frac{1}{p}x-\frac{r}{p}\Big),
\quad x\in \bQ_p,
$$
whose solution ({\it a refinable function}) $\phi$ is the
characteristic function $\Omega\big(|x|_p\big)$ of the unit disc.
The conjecture to use the above equation as a {\it refinement
equation} was proposed in~\cite{Kh-Sh1}. The above {\it refinement
equation} is {\it natural} and reflects the fact that the
characteristic function of the unit disc $B_{0}=\{x: |x|_p \le 1\}$
is represented as a sum of $p$  characteristic functions of
the mutually disjoint discs
$B_{-1}(r)=\big\{x: \big|\frac{x}{p}-\frac{r}{p}\big|_p \le 1\big\}$,
$r=0,1,\dots,p-1$ (see (\ref{79})).
In Subsec.~\ref{s4.3}, the $2$-adic Haar MRA is constructed.
Namely, we proved that the {\it refinable function}
$\phi(x)=\Omega\big(|x|_2\big)$ generates a MRA, which is an
analog of the classical Haar MRA.
It is shown that a $2$-adic analog of the real  wavelet function generated by
Haar MRA   generates an orthonormal basis (\ref{62.0-7})
for ${\cL}^2(\bQ_2)$. This basis coincides with Kozyrev's one (\ref{62.0-1}) for
 $p=2$. We proved that  Kozyrev's  basis is not  a unique orthonormal
wavelet basis generated by $2$-adic Haar MRA.

In Sec.~\ref{s5}, {\it infinity many  $2$-adic  Haar
wavelet  functions} generating different orthonormal bases for ${\cL}^2(\bQ_2)$ are found.
Explicit  formulas for  Haar wavelet
functions $\psi^{(s)}$, $s\in \bN$, (Theorem~\ref{th4}) are given in Subsec.~\ref{s5.1}.
All real wavelet functions are described for
$s=1,2$ in Subsec.~\ref{s5.2}.

In Sec.~\ref{s6},  we study multivariate Haar bases.
A  general scheme for construction of $2$-adic multidimensional separable MRA
is described in Subsec.~\ref{s6.1},.
 According to this scheme, separable $2$-adic
Haar wavelets  (\ref{62.9}) in ${\cL}^2(\bQ_2^n)$
are constructed in Subsec.~\ref{s6.2}.

Sec.~\ref{s8} is devoted to the spectral theory of pseudo-differential
operators (\ref{64.3}). Theorems~\ref{th5},~\ref{th6} state
a criterion for pseudo-differential operators (\ref{64.3}) to
have multidimensional wavelets (\ref{w-62.8=1}) or (\ref{62.9})
as eigenfunctions. In particular, according to
Corollaries~\ref{cor5},~\ref{cor6}, the multidimensional
wavelets (\ref{w-62.8=1}) and (\ref{62.9}) are eigenfunctions
of the multidimensional fractional operator (\ref{59**}).

\section{Preliminary results in $p$-adics}
\label{s2}

\subsection{$p$-Adic distributions.}\label{s2.1}
Here and in what follows, we shall systematically use the
notations and the results from~\cite{Vl-V-Z} and~\cite[Ch.II]{G-Gr-P}.
Let $\bN$, $\bZ$, $\bC$ be the sets of positive integers, integers,
complex numbers, respectively,  $\bN_0:=\{0\}\cup\bN$.

The field of $p$-adic numbers is denoted by $\bQ_p$.
The canonical form of any $p$-adic number $x\ne 0$ is
\begin{equation}
\label{2}
x=p^{\gamma}(x_0 + x_1p + x_2p^2 + \cdots),
\end{equation}
where $\gamma=\gamma(x)\in \bZ$, \ $x_j=0,1,\dots,p-1$, $x_0\ne 0$,
$j=0,1,\dots$. The series is convergent in the $p$-adic norm (\ref{1}),
and one has $|x|_p=p^{-\gamma}$.
By means of representation (\ref{2}), the {\it fractional part} $\{x\}_p$
of a number $x\in \bQ_p$ is defined as follows
\begin{equation}
\label{8.2**}
\{x\}_p=\left\{
\begin{array}{lll}
0,\quad \text{if} \quad \gamma(x)\ge 0 \quad  \text{or} \quad x=0,&&  \\
p^{\gamma}(x_0+x_1p+x_2p^2+\cdots+x_{|\gamma|-1}p^{|\gamma|-1}),
\quad \text{if} \quad \gamma(x)<0. && \\
\end{array}
\right.
\end{equation}

The function
\begin{equation}
\label{8.2-1**}
\chi_p(\xi x)=e^{2\pi i\{\xi x\}_p}
\end{equation}
for every fixed $\xi \in \bQ_p$ is an {\it additive character} of
the field $\bQ_p$, where $\{\cdot\}_p$ is a fractional part (\ref{8.2**}).

The space $\bQ_p^n:=\bQ_p\times\cdots\times\bQ_p$ consists of points
$x=(x_1,\dots,x_n)$, where $x_j \in \bQ_p$, $j=1,2\dots,n$, \ $n\ge 2$.
The $p$-adic norm on $\bQ_p^n$ is
\begin{equation}
\label{8}
|x|_p=\max_{1 \le j \le n}|x_j|_p, \quad x\in \bQ_p^n,
\end{equation}
where $|x_j|_p$ is defined by (\ref{1}).

Denote by $B_{\gamma}^n(a)=\{x\in \bQ_p^n: |x-a|_p \le p^{\gamma}\}$
the ball of radius $p^{\gamma}$ with the center at a point
$a=(a_1,\dots,a_n)\in \bQ_p^n$ and by
$S_{\gamma}^n(a)=\{x\in \bQ_p^n: |x-a|_p = p^{\gamma}\}
=B_{\gamma}^n(a)\setminus B_{\gamma-1}^n(a)$ its boundary (sphere),
$\gamma \in \bZ$. For $a=0$, we set $B_{\gamma}^n(0)=B_{\gamma}^n$ and
$S_{\gamma}^n(0)=S_{\gamma}^n$. For the case $n=1$, we will omit the
upper index $n$.
It is clear that
\begin{equation}
\label{9}
B_{\gamma}^n(a)=B_{\gamma}(a_1)\times\cdots\times B_{\gamma}(a_n),
\end{equation}
where $B_{\gamma}(a_j)=\{x_j: |x_j-a_j|_p \le p^{\gamma}\}\subset\bQ_p$
is the disc (one-dimensional ball) of radius $p^{\gamma}$ with the center at a point $a_j\in \bQ_p$,
$j=1,2\dots,n$.

Any two balls in $\bQ_p^n$ either are disjoint or one
contains the other. Every point of a ball is its center.

According to~\cite[I.3,Examples 1,2.]{Vl-V-Z}, a one-dimensional
disc $B_{\gamma}$ is represented as a sum of $p^{\gamma-\gamma'}$
mutually {\it disjoint} discs $B_{\gamma'}(a)$, $\gamma'<\gamma$:
\begin{equation}
\label{79.0}
B_{\gamma}=B_{\gamma'}\cup\Big(\cup_{a}B_{\gamma'}(a)\Big),
\end{equation}
where $a=0$ and
$a=a_{-r}p^{-r}+a_{-r+1}p^{-r+1}+\cdots+a_{-\gamma'-1}p^{-\gamma'-1}$
are the centers of the discs $B_{\gamma'}(a)$, \
$r=\gamma,\gamma-1,\gamma-2,\dots,\gamma'+1$, \, $0\le a_j\le p-1$,
\, $a_{-r}\ne 0$.
In particular, the disc $B_{0}$ is represented as a sum of $p$
mutually {\it disjoint} discs
\begin{equation}
\label{79}
B_{0}=B_{-1}\cup\Big(\cup_{r=1}^{p-1}B_{-1}(r)\Big),
\end{equation}
where $B_{-1}(r)=\{x\in S_{0}: x_0=r\}=r+p\bZ_p$, $r=1,\dots,p-1$;
$B_{-1}=\{|x|_p\le p^{-1}\}=p\bZ_p$; and
$S_{0}=\{|x|_p=1\}=\cup_{r=1}^{p-1}B_{-1}(r)$. Here all the discs
are mutually disjoint. Coverings (\ref{79.0}) and (\ref{79}) are called the
{\it canonical covering} of the discs $B_{0}$ and $B_{\gamma}$,
respectively.

There exists the Haar measure $dx$ on $\bQ_p$. This  measure is  positive,
invariant under the shifts, i.e., $d(x+a)=dx$, and normalized by
$\int_{|\xi|_p\le 1}\,dx=1$.
The invariant measure $dx$ on the field $\bQ_p$ is extended to an
invariant measure $d^n x=dx_1\cdots dx_n$ on $\bQ_p^n$ in the standard way.

A complex-valued function $f$ defined on $\bQ_p^n$ is called
{\it locally-constant} if for any $x\in \bQ_p^n$ there exists
an integer $l(x)\in \bZ$ such that
$$
f(x+y)=f(x), \quad y\in B_{l(x)}^n.
$$

Let ${\cE}(\bQ_p^n)$ and ${\cD}(\bQ_p^n)$ be the linear spaces of
locally-constant $\bC$-valued functions on $\bQ_p^n$ and locally-constant
$\bC$-valued functions with compact supports (so-called test functions),
respectively~\cite[VI.1.,2.]{Vl-V-Z}. If $\varphi \in {\cD}(\bQ_p^n)$,
according to Lemma~1 from~\cite[VI.1.]{Vl-V-Z}, there exists $l\in \bZ$,
such that
$$
\varphi(x+y)=\varphi(x), \quad y\in B_l^n, \quad x\in \bQ_p^n.
$$
The largest of such numbers $l=l(\varphi)$ is called the
{\it parameter of constancy} of the function $\varphi$.
Let us denote by ${\cD}^l_N(\bQ_p^n)$ the finite-dimensional space of
test functions from ${\cD}(\bQ_p^n)$ having supports in the ball $B_N^n$
and with parameters of constancy $\ge l$~\cite[VI.2.]{Vl-V-Z}.
The following embedding holds:
${\cD}^l_N(\bQ_p^n)\subset {\cD}^{l'}_{N'}(\bQ_p^n)$, \ $N\le N'$,
$l\ge l'$. Thus ${\cD}(\bQ_p^n)
=\lim{\rm ind}_{N\to \infty}\lim{\rm ind}_{l\to -\infty}{\cD}^l_N(\bQ_p^n)$.
The space ${\cD}(\bQ_p^n)$ is a complete locally convex vector space.
According to~\cite[VI,(5.2')]{Vl-V-Z}, any function
$\varphi \in {\cD}^l_N(\bQ_p^n)$ is represented in the
following form
\begin{equation}
\label{9.4}
\varphi(x)=\sum_{\nu=1}^{p^{n(N-l)}}\varphi(c^{\nu})
\Omega(p^{-l}|x-c^{\nu}|_p), \quad x\in \bQ_p^n,
\end{equation}
where $\Omega(p^{-l}|x-c^{\nu}|_p)$ are the characteristic
functions of the mutually disjoint balls $B_{l}(c^{\nu})$, and
the points $c^{\nu}=(c_1^{\nu},\dots c_n^{\nu})\in B_N^n$
do not depend on $\varphi$.

Denote by ${\cD}'(\bQ_p^n)$ the set of all linear functionals
($p$-adic distributions) on ${\cD}(\bQ_p^n)$~\cite[VI.3.]{Vl-V-Z}.

Let us introduce in ${\cD}(\bQ_p^n)$ a {\it canonical
$\delta$-sequence} $\delta_k(x)=p^{nk}\Omega(p^k|x|_p)$,
and a {\it canonical $1$-sequence} $\Omega(p^{-k}|x|_p)$
(the characteristic function of the ball $B_{k}^n$), $k \in \bZ$, \
$x\in \bQ_p^n$, where
\begin{equation}
\label{10}
\Omega(t)=\left\{
\begin{array}{lll}
1, && 0 \le t \le 1, \\
0, && t>1, \,\, t\in \bR. \\
\end{array}
\right.
\end{equation}
It is clear that $\delta_k \to \delta$, $k \to \infty$ in
${\cD}'(\bQ_p^n)$, where $\delta$ is is the Dirac delta function,
and $\Omega(p^{-k}|\cdot|_p)\to 1$, $k \to \infty$ in
${\cE}(\bQ_p^n)$ (see~\cite[VI.3., VII.1.]{Vl-V-Z}).

The Fourier transform of $\varphi\in {\cD}(\bQ_p^n)$ is defined by the
formula
$$
F[\varphi](\xi)=\int_{\bQ_p^n}\chi_p(\xi\cdot x)\varphi(x)\,d^nx,
\quad \xi \in \bQ_p^n,
$$
where $\chi_p(\xi\cdot x)=\chi_p(\xi_1x_1)\cdots\chi_p(\xi_nx_n)
=e^{2\pi i\sum_{j=1}^{n}\{\xi_j x_j\}_p}$; \
$\xi\cdot x$ is the scalar product of vectors and $\chi_p(\xi_jx_j)$
are additive characters (\ref{8.2-1**}).
The Fourier transform is a linear isomorphism ${\cD}(\bQ_p^n)$ into
${\cD}(\bQ_p^n)$. Moreover, according
to~\cite[Lemma~A.]{Taib1},~\cite[III,(3.2)]{Taib3},
~\cite[VII.2.]{Vl-V-Z},
\begin{equation}
\label{12}
\varphi(x) \in {\cD}^l_N(\bQ_p^n) \quad \text{iff} \quad
F\big[\varphi(x)\big](\xi) \in {\cD}^{-N}_{-l}(\bQ_p^n).
\end{equation}
The Fourier transform $F[f]$ of a distribution
$f\in {\cD}'(\bQ_p^n)$ is defined by
$$
\langle F[f],\varphi\rangle=\langle f,F[\varphi]\rangle,
\quad \forall \, \varphi\in {\cD}(\bQ_p^n).
$$

Let $A$ be a matrix and $b\in \bQ_p^n$. Then for a distribution
$f\in{\cD}'(\bQ_p^n)$ the following relation holds~\cite[VII,(3.3)]{Vl-V-Z}:
\begin{equation}
\label{14}
F[f(Ax+b)](\xi)
=|\det{A}|_p^{-1}\chi_p\big(-A^{-1}b\cdot \xi\big)F[f(x)]\big(A^{-1}\xi\big),
\end{equation}
where $\det{A} \ne 0$.
According to~\cite[IV,(3.1)]{Vl-V-Z},
\begin{equation}
\label{14.1}
F[\Omega(p^{-k}|\cdot|_p)](x)=\delta_{k}(x), \quad k\in \bZ,
\quad x \in \bQ_p^n.
\end{equation}
In particular, $F[\Omega(|\cdot|_p)](x)=\Omega(|x|_p)$.

\subsection{The $p$-adic Lizorkin spaces.}\label{s2.2}
Let us introduce a space of the $p$-adic {\it Lizorkin test functions\/}
(see~\cite{Al-Kh-Sh3},~\cite{Al-Kh-Sh4})
$$
\Phi(\bQ_p^n)=\{\phi: \phi=F[\psi], \, \psi\in \Psi(\bQ_p^n)\},
$$
where
$\Psi(\bQ_p^n)=\{\psi(\xi)\in \cD(\bQ_p^n): \psi(0)=0\}$.
Here $\Psi(\bQ_p^n), \Phi(\bQ_p^n)\subset \cD(\bQ_p^n)$.
The space $\Phi(\bQ_p^n)$ can be equipped with the
topology of the space $\cD(\bQ_p^n)$ which makes $\Phi$ a
complete space.

In view of (\ref{12}),
\begin{equation}
\label{54}
\phi\in \Phi(\bQ_p^n) \, \Longleftrightarrow \,
\int_{\bQ_p^n}\phi(x)\,d^nx=0, \quad \phi\in \cD(\bQ_p^n).
\end{equation}
Moreover, $\phi \in {\cD}^l_N(\bQ_p^n)\cap\Phi(\bQ_p^n)$, i.e.,
$\int_{B^n_{N}}\phi(x)\,d^nx=0$, iff \
$\psi=F^{-1}[\phi]\in {\cD}^{-N}_{-l}(\bQ_p^n)\cap\Psi(\bQ_p^n)$,
i.e., $\psi(\xi)=0$, $\xi \in B^n_{-N}$.

Let $\Phi'(\bQ_p^n)$ and $\Psi'(\bQ_p^n)$ denote the topological
dual of the spaces $\Phi(\bQ_p^n)$ and $\Psi(\bQ_p^n)$, respectively.
We call $\Phi'(\bQ_p^n)$ the space of  $p$-adic {\it Lizorkin
distributions\/}.

If $\Psi^{\perp}$ and $\Phi^{\perp}$ are subspaces of
functionals in $\cD'(\bQ_p^n)$ orthogonal to $\Psi(\bQ_p^n)$
and $\Phi(\bQ_p^n)$, respectively, then
$\Psi^{\perp}=\{f\in \cD'(\bQ_p^n): f=C\delta, \, C\in \bC\}$
and $\Phi^{\perp}=\{f\in \cD'(\bQ_p^n): f=C, \, C\in \bC\}$.
According to~\cite{Al-Kh-Sh3}, $\Phi'(\bQ_p^n)=\cD'(\bQ_p^n)/\Phi^{\perp}$
and $\Psi'(\bQ_p^n)=\cD'(\bQ_p^n)/\Psi^{\perp}$.
Thus the space $\Phi'(\bQ_p^n)$ can be obtained from $\cD'(\bQ_p^n)$
by ``sifting out'' constants.

We define the Fourier transform of
$f\in \Phi'(\bQ_p^n)$ and $g\in \Psi'(\bQ_p^n)$
respectively by
\begin{equation}
\label{51}
\begin{array}{rcl}
\displaystyle
\langle F[f],\psi\rangle=\langle f,F[\psi]\rangle,
&& \forall \, \psi\in \Psi(\bQ_p^n), \medskip \\
\displaystyle
\langle F[g],\phi\rangle=\langle g,F[\phi]\rangle,
&& \forall \, \phi\in \Phi(\bQ_p^n). \\
\end{array}
\end{equation}
Since $F[\Phi(\bQ_p^n)]=\Psi(\bQ_p^n)$ and
$F[\Psi(\bQ_p^n)]=\Phi(\bQ_p^n)$, formulas (\ref{51})
give well defined objects.
It is clear that $F[\Phi'(\bQ_p^n)]=\Psi'(\bQ_p^n)$
and $F[\Psi'(\bQ_p^n)]=\Phi'(\bQ_p^n)$~\cite{Al-Kh-Sh3}.

Recall that the {\it usual} Lizorkin spaces were studied in the
excellent papers of P.~I.~Lizorkin~\cite{Liz1},~\cite{Liz3}
(see also~\cite{Sam3},~\cite{Sam-Kil-Mar}).

\subsection{Pseudo-differential operators in the Lizorkin space.}\label{s7.1}
Consider the following class of pseudo-differential operators $A$ in the Lizorkin
space of the test functions $\Phi(\bQ_p^n)$ defined by
$$
(A\phi)(x)=F^{-1}\big[\cA\,F[\phi]\big](x)
\qquad\qquad\qquad\qquad\qquad\qquad\qquad
$$
\begin{equation}
\label{64.3}
=\int_{\bQ_p^n}\int_{\bQ_p^n}\chi_p\big((y-x)\cdot \xi\big)
\cA(\xi)\phi(y)\,d^n\xi\,d^ny,
\quad \phi \in \Phi(\bQ_p^n),
\end{equation}
with symbols $\cA\in \cE(\bQ_p^n\setminus \{0\})$, which were
introduced in~\cite{Al-Kh-Sh3}.

\begin{Lemma}
\label{lem4}
{\rm (~\cite{Al-Kh-Sh3})}
The Lizorkin space $\Phi(\bQ_p^n)$ is invariant under the
pseudo- differential operators {\rm(\ref{64.3})}. Moreover,
$A(\Phi(\bQ_p^n))=\Phi(\bQ_p^n)$.
\end{Lemma}

Given pseudo-differential $A$ with a symbols $\cA$,
 define the conjugate  operator $A^{T}$ by
\begin{equation}
\label{64.5}
(A^{T}\phi)(x)=F^{-1}[\cA(-\cdot)F[\phi]](x)
=\int_{\bQ_p^n}\chi_p(-x\cdot \xi)\cA(-\xi)F[\phi](\xi)\,d^n\xi
\end{equation}
and introduce the corresponding
 operator $A$ in the space of Lizorkin
distributions: if $f \in \Phi'(\bQ_p^n)$, then
\begin{equation}
\label{64.4}
\langle Af,\phi\rangle=\langle f,A^{T}\phi\rangle,
\qquad \forall \, \phi \in \Phi(\bQ_p^n).
\end{equation}
It is clear that
\begin{equation}
\label{64.3*}
Af=F^{-1}[\cA\,F[f]]\in \Phi'(\bQ_p^n),
\end{equation}
i.e., the  space of Lizorkin distributions $\Phi'(\bQ_p^n)$
is invariant under pseudo-differential operators $A$.
Moreover, by Lemma~\ref{lem4}, $A(\Phi'(\bQ_p^n))=\Phi'(\bQ_p^n)$.

If $A, B$ are pseudo-differential operators with symbols
$\cA, \cB\in \cE(\bQ_p^n\setminus \{0\})$ respectively,
then the operator $AB$ is well defined and represented by the formula
$$
(AB)f=F^{-1}[\cA\cB\,F[f]]\in \Phi'(\bQ_p^n).
$$
If $\cA(\xi)\ne 0$ for all $\xi\in \bQ_p^n\setminus \{0\}$, then
operator
we
the  pseudo-differential operator
$$
A^{-1}f=F^{-1}[\cA^{-1}\,F[f]], \quad f\in \Phi'(\bQ_p^n).
$$
is, evidently, the inverse operator to $A$.

Thus the family of pseudo-differential operators $A$ whose symbols $\cA$
do not vanish on $\bQ_p^n\setminus \{0\}$, forms an Abelian group.

\subsection{Multidimensional fractional operator.}\label{s7.2}
Let us consider a pseudo- differential operator $D^{\alpha}$
with the symbol $\cA(\xi)=|\xi|_p^{\alpha}$, $\alpha\in \bC$,
$\alpha\ne -n$. By definition~(\ref{64.3}),
\begin{equation}
\label{61**}
\big(D^{\alpha}\phi\big)(x)
=F^{-1}\big[|\cdot|^{\alpha}_pF[\phi](\cdot)\big](x),
\quad \phi \in \Phi(\bQ_p^n).
\end{equation}
These operators were introduced by Taibleson~\cite[\S2]{Taib1},~\cite[III.4.]{Taib3}
as operators in the space of distributions ${\cD}'(\bQ_p^n)$.
The relation (\ref{61**}) can be rewritten as the following convolution
(see~\cite[VII.1.]{Vl-V-Z}):
$\big(D^{\alpha}\phi\big)(x)\stackrel{def}{=}\kappa_{-\alpha}(x)*\phi(x)
=\langle \kappa_{-\alpha}(\xi),\phi(x-\xi)\rangle$,
\ $x\in \bQ_p^n$, $\phi\in \Phi(\bQ_p^n)$,
where the distribution (from ${\cD}'(\bQ_p^n)$)
\begin{equation}
\label{63.4}
\kappa_{\alpha}(x)=\frac{|x|_p^{\alpha-n}}{\Gamma_p^{(n)}(\alpha)},
\quad \alpha \ne 0, \, \,  n, \qquad x\in \bQ_p^n,
\end{equation}
is called the multidimensional {\it Riesz kernel\/},
$\Gamma_p^{(n)}(\alpha)$ is the $n$-dimensional
$\Gamma$-{\it function\/} (see for details, see ~\cite{Taib1},
~\cite[III]{Taib3},~\cite[VIII]{Vl-V-Z}).

The Riesz kernel has a removable singularity at $\alpha=0$ and according
to~\cite[\S2]{Taib1},~\cite[III.4.]{Taib3},~\cite[VIII.2]{Vl-V-Z},
we have
\begin{equation}
\label{63.5}
\kappa_{0}(x)\stackrel{def}{=}\lim_{\alpha\to 0}\kappa_{\alpha}(x)=\delta(x).
\end{equation}
According to~\cite{Al-Kh-Sh3},~\cite{Al-Kh-Sh4}, one can similarly introduce
a  {\it Lizorkin  distribution\/} $\kappa_{n}(\cdot)$ by
\begin{equation}
\label{63.7}
\kappa_{n}(x)\stackrel{def}{=}\lim_{\alpha \to n}\kappa_{\alpha}(x)
=-\frac{1-p^{-n}}{\log p}\log|x|_p.
\end{equation}

Due to (\ref{63.5}), (\ref{63.7}), we can define the operator (\ref{61**})
in the Lizorkin space of test functions $\phi\in \Phi(\bQ_p^n)$ by
\begin{equation}
\label{59**}
\big(D^{\alpha}\phi\big)(x)\stackrel{def}{=}\kappa_{-\alpha}(x)*\phi(x)
=\langle \kappa_{-\alpha}(\xi),\phi(x-\xi)\rangle, \quad x\in \bQ_p^n,
\quad \alpha \in \bC.
\end{equation}
According to Lemma~\ref{lem4}, the Lizorkin space $\Phi(\bQ_p^n)$ is
invariant under the Taibleson fractional operator $D^{\alpha}$
and $D^{\alpha}(\Phi(\bQ_p^n))=\Phi(\bQ_p^n)$.
By (\ref{64.5}), (\ref{64.4}), $(D^{\alpha})^{T}=D^{\alpha}$
and for any $f\in \Phi'(\bQ_p^n)$ we have
\begin{equation}
\label{62**}
\langle D^{\alpha}f,\phi\rangle\stackrel{def}{=}
\langle f, D^{\alpha}\phi\rangle,
\quad \forall \, \phi\in \Phi(\bQ_p^n).
\end{equation}
It is clear that $D^{\alpha}(\Phi'(\bQ_p^n))=\Phi'(\bQ_p^n)$.

It is easy to prove that
$\kappa_{\alpha}(x)*\kappa_{\beta}(x)=\kappa_{\alpha+\beta}(x)$,
$\alpha, \beta \in \bC$, holds in the sense of the Lizorkin space
$\Phi'(\bQ_p^n)$ (see~\cite{Al-Kh-Sh3}). Consequently, the family of
operators $D^{\alpha}$, $\alpha \in \bC$, in the Lizorkin space
forms an Abelian group: if $f \in \Phi'(\bQ_p^n)$ then
$D^{\alpha}D^{\beta}_{x}f=
D^{\beta}D^{\alpha}f=D^{\alpha+\beta}f$, \
$D^{\alpha}D^{-\alpha}f=f$, \ $\alpha,\beta \in \bC$.

\section{Multiresolution analysis (one-dimensional case)}
\label{s4}

\subsection{$p$-Adic multiresolution analysis.}\label{s4.1}
Consider the set
$$
I_p=\{a=p^{-\gamma}\big(a_{0}+a_{1}p+\cdots+a_{\gamma-1}p^{\gamma-1}\big):
\qquad\qquad\qquad\qquad
$$
\begin{equation}
\label{62.0**}
\qquad\qquad
\gamma\in \bN; \, a_j=0,1,\dots,p-1; \, j=0,1,\dots,\gamma-1\}.
\end{equation}
This set can be identified with the factor group $\bQ_p/\bZ_p$.

It is well known that
$\bQ_p=B_{0}\cup\cup_{\gamma=1}^{\infty}S_{\gamma}$, where
$S_{\gamma}=\{x\in \bQ_p: |x|_p = p^{\gamma}\}$. Due to
(\ref{2}), $x\in S_{\gamma}$, $\gamma\ge 1$, if and only if
$x=x_{-\gamma}p^{-\gamma}+x_{-\gamma+1}p^{-\gamma+1}+\cdots+x_{-1}p^{-1}+\xi$,
where $x_{-\gamma}\ne 0$, $\xi \in B_{0}$. Since
$x_{-\gamma}p^{-\gamma}+x_{-\gamma+1}p^{-\gamma+1}
+\cdots+x_{-1}p^{-1}\in I_p$, we have a ``natural'' decomposition of
$\bQ_p$ to a union of mutually  disjoint discs:
$$
\bQ_p=\bigcup\limits_{a\in I_p}B_{0}(a).
$$
So, $I_p$ is a ``natural'' group of shifts for $\bQ_p$, which will
be used in the sequel.

\begin{Definition}
\label{de1} \rm
A collection of closed spaces
$V_j\subset\cL^2(\bQ_p)$, $j\in\bZ$, is called a
{\it multiresolution analysis {\rm(}MRA{\rm)} in $\cL^2(\bQ_p)$} if the
following axioms hold

(a) $V_j\subset V_{j+1}$ for all $j\in\bZ$;

(b) $\bigcup\limits_{j\in\bZ}V_j$ is dense in $\cL^2(\bQ_p)$;

(c) $\bigcap\limits_{j\in\bZ}V_j=\{0\}$;

(d) $f(\cdot)\in V_j \Longleftrightarrow f(p^{-1}\cdot)\in V_{j+1}$
for all $j\in\bZ$;

(e) there exists a function $\phi \in V_0$
such that the system $\{\phi(\cdot-a), a\in I_p\}$ is an orthonormal
basis for $V_0$.
\end{Definition}

The function $\phi$ from axiom (e) is called
{\em refinable}  or {\em scaling}. It follows  immediately
from axioms (d) and (e) that the functions $p^{j/2}\phi(p^{-j}\cdot-a)$,
$a\in I_p$, form an orthonormal basis for $V_j$, $j\in\bZ$.

According to the standard scheme (see, e.g.,~\cite[\S 1.3]{NPS})
for construction of MRA-based wavelets, for each $j$, we define
a space $W_j$ ({\em wavelet space}) as the orthogonal complement
of $V_j$ in $V_{j+1}$, i.e.,
\begin{equation}
\label{61}
V_{j+1}=V_j\oplus W_j, \qquad j\in \bZ,
\end{equation}
where $W_j\perp V_j$, $j\in \bZ$. It is not difficult to see that
\begin{equation}
\label{61.0}
f\in W_j \Longleftrightarrow f(p^{-1}\cdot)\in W_{j+1},
\quad\text{for all}\quad j\in \bZ
\end{equation}
and $W_j\perp W_k$, $j\ne k$.
Taking into account axioms (b) and (c), we obtain
\begin{equation}
\label{61.1}
{\bigoplus\limits_{j\in\bZ}W_j}=\cL^2(\bQ_p)
\quad \text{(orthogonal direct sum)}.
\end{equation}

If now we find  a function $\psi \in W_0$
such that the system $\{\psi(x-a), a\in~I_p\}$ is an orthonormal
basis for $W_0$, then, due to~(\ref{61.0}) and (\ref{61.1}),
the system $\{p^{j/2}\psi(p^{-j}\cdot-a), a\in I_p, j\in\bZ\}$,
is an orthonormal basis for $\cL^2(\bQ_p)$.
Such a function $\psi$ is called a {\em wavelet function} and
the basis is a {\em wavelet basis}.

\subsection{$p$-Adic refinement equation.}\label{s4.2}
Let $\phi$ be a refinable function for a MRA. As was mentioned above,
the system $\{p^{1/2}\phi(p^{-1}\cdot-a), a\in I_p\}$,
is a basis for $V_1$. It follows from axoim (a) that
\begin{equation}
\label{62.0-2*}
\phi=\sum_{a\in I_p}\alpha_a\phi(p^{-1}\cdot-a),
\quad \alpha_a\in \bC.
\end{equation}
We see that the function $\phi$ is a solution of a
special kind of functional equation. Such equations are called
{\em refinement equations}. Investigation of refinement equations and
their solutions is the most difficult part of the wavelet theory in
real analysis.

A natural way for construction of a MRA (see, e.g.,~\cite[\S 1.2]{NPS})
is the following. We start with an appropriate function $\phi$
whose integer shifts form an orthonormal system and set
$V_j=\overline{{\rm span}\big\{\phi\big(p^{-j}\cdot-a\big):a\in I_p\big\}}$,
\ $j\in \bZ$.
It is clear that axioms (d) and (e) of Definition~\ref{de1} are fulfilled.
Of course, not any such a function $\phi$ provides axiom $(a)$.
In the {\em real setting}, the relation $V_0\subset V_{1}$ holds
if and only if the refinable function satisfies a refinement equation.
Situation is different in $p$-adics. Generally speaking, a refinement
equation (\ref{62.0-2*}) {\em does not imply the including property}
$V_0\subset V_{1}$. Indeed, we need all the functions $\phi(\cdot-b)$,
$b\in I_p$, to belong to the space $V_1$, i.e., the identities
$\phi(x-b)=\sum_{a\in I_p}\alpha_{a,b}\phi(p^{-1}x-a)$ should be
fulfilled for all $b\in I_p$. Since $p^{-1}b+a$ is not in $I_p$ in general,
we {\em can not state} that
$\phi(x-b)=\sum_{a\in I_p}\alpha_{a,b}\phi(p^{-1}x-p^{-1}b-a)\in V_1$
for all $b\in I_p$. Nevertheless, some refinable equations
imply including imply property, which may happen because of different causes.

The {\it refinement equation} reflects {\it some
``self-similarity''}. The structure of the space $\bQ_p$ has
a {\it natural} ``self-similarity'' property which is given by
formulas (\ref{79.0}), (\ref{79}). By
(\ref{79}), the characteristic function
$\Omega\big(|x|_p\big)$ of the unit disc $B_{0}$ is
represented as a sum of $p$  characteristic functions of the
mutually disjoint discs $B_{-1}(r)$, $r=0,1,\dots,p-1$, i.e.,
\begin{equation}
\label{62.0-2}
\Omega\big(|x|_p\big)=\sum_{r=0}^{p-1}\Omega\big(p|x-r|_p\big)=
\sum_{r=0}^{p-1}\Omega\Big(\Big|\frac{1}{p}x-\frac{r}{p}\Big|_p\Big),
\quad x\in \bQ_p.
\end{equation}
Thus, in $p$-adics, we have a {\it natural} {\it refinement equation}
(\ref{62.0-2*}):
\begin{equation}
\label{62.0-3}
\phi(x)=\sum_{r=0}^{p-1}\phi\Big(\frac{1}{p}x-\frac{r}{p}\Big),
\quad x\in \bQ_p,
\end{equation}
whose solution is $\phi(x)=\Omega\big(|x|_p\big)$.
This equation is an analog of the {\em refinement equation}
generating the Haar MRA in real analysis.

\subsection{Construction of $2$-adic Haar multiresolution analysis.}\label{s4.3}
Using the {\it refinement equation} (\ref{62.0-3}) for $p=2$
\begin{equation}
\label{62.0-4}
\phi(x)=\phi\Big(\frac{1}{2}x\Big)+\phi\Big(\frac{1}{2}x-\frac{1}{2}\Big),
\quad x\in \bQ_2,
\end{equation} and its solution, the {\it refinable function}
$\phi(x)=\Omega\big(|x|_2\big)$,
we construct a $2$-adic multiresolution analysis.

Set
\begin{equation}
\label{70.1}
V_j=\overline{{\rm span}\big\{\phi\big(2^{-j}x-a\big):a\in I_2\big\}},
\quad j\in \bZ.
\end{equation}
It is clear that axioms (d) and (e) of Definition~\ref{de1} are
fulfilled and the system $\{2^{j/2}\phi(2^{-j}\cdot-a), a\in I_p\}$
is an orthonormal basis for $V_j$, $j\in \bZ$.
Since the numbers $2^{-1}b$, $2^{-1}b+2^{-1}$
are in $I_2$ for all $b\in I_2$, it follows from the refinement
equation (\ref{62.0-4}) that $V_0\subset V_1$. By the definition
(\ref{70.1}) of the spaces $V_j$, this yields axiom $(a)$. Due
to the {\it refinement equation} (\ref{62.0-4}), we obtain
that $V_j\subset V_{j+1}$, i.e., the axiom (a) from
Definition~\ref{de1} holds.

Note that the characteristic function of the unit disc
$\Omega\big(|x|_2\big)$ has a wonderful feature:
$\Omega(|\cdot+\xi|_2)=\Omega(|\cdot|_2)$, for all
$\xi\in \bZ_2$ because the $p$-adic norm is non-Archimedean.
In particular, $\Omega(|\cdot\pm 1|_2)=\Omega(|\cdot|_2)$, i.e.,
\begin{equation}
\label{62.0-0}
\phi(x\pm 1)=\phi(x), \quad \forall \, x\in \bQ_2.
\end{equation}
Thus $\phi$ is a {\em $1$-periodic} function.

\begin{Proposition}
\label{pr2}
The axiom $(b)$ of Definition~{\rm\ref{de1}} holds, i.e., \,
$\overline{\cup_{j\in\bZ}V_j}=\cL^2(\bQ_2)$.
\end{Proposition}

\begin{proof}
According to (\ref{9.4}), any function $\varphi\in \cD(\bQ_2)$
belongs to one of the spaces ${\cD}^l_N(\bQ_2)$, and consequently,
is represented in the form
\begin{equation}
\label{71}
\varphi(x)=\sum_{\nu=1}^{2^{N-l}}
\varphi(c^{\nu})\Omega(2^{-l}|x-c^{\nu}|_2), \quad x\in \bQ_2,
\end{equation}
where
$c^{\nu}\in B_{N}$, \ $\nu=1,2,\dots 2^{N-l}$; \ $l=l(\varphi)$,
$N=N(\varphi)$; \ $l\in \bZ$. Taking into account that
$\Omega(2^{-l}|x-c^{\nu}|_2)=\Omega(|2^{l}x-2^{l}c^{\nu}|_2)
=\phi(2^{l}x-2^{l}c^{\nu})$, we can rewrite~(\ref{71}) as
$$
\varphi(x)=\sum_{\nu=1}^{2^{N-l}}\alpha_{\nu}\phi(2^{l}x-2^{l}c^{\nu})
\quad x\in \bQ_2, \quad c^{\nu}\in B_{N}, \quad \alpha_{\nu}\in \bC.
$$
Since any number $2^{l}c^{\nu}$
can be represented in the form $2^{l}c^{\nu}=a^{\nu}+b^{\nu}$,
$a^{\nu}\in I_2$, $b^{\nu}\in \bZ_2$, using~(\ref{62.0-0}),
we have
$$
\varphi(x)=\sum_{\nu=1}^{2^{N-l}}\alpha_{\nu}
\phi(2^lx-a^{\nu}), \quad x\in \bQ_2, \quad a^{\nu}\in I_2,
\quad \alpha_{\nu}\in \bC,
$$
i.e., $\varphi(x)\in V_{-l}$. Thus any test function
$\varphi$ belongs to one of the space $V_{j}$, where
$j=j(\varphi)$, \ $j\in \bZ$.

Since the space $\cD(\bQ_2)$ is dense in
$\cL^{2}(\bQ_2)$~\cite[VI.2]{Vl-V-Z}, approximating
any function from $\cL^{2}(\bQ_2)$ by test functions
$\varphi\in \cD(\bQ_2)$, we prove our assertion.
\end{proof}

\begin{Proposition}
\label{pr3}
The axiom $(c)$ of Definition~{\rm\ref{de1}} holds, i.e.,
$\cap_{j\in\bZ}V_j=\{0\}$.
\end{Proposition}

\begin{proof}
Assume that $\cap_{j\in\bZ}V_j\ne\{0\}$. Then there exists a
function $f\in \cD(\bQ_2)$ such that $\|f\|\ne0$ and
 $f\in V_j$ for all $j\in\bZ$. Hence, due to (\ref{70.1}), we have
$f(x)=\sum_{a\in I_2}c_{ja}\phi\big(2^{-j}x-a\big)$ for all $j\in\bZ$.

Let $x=2^{-N}(x_{0}+x_{1}2+x_{2}2^2+\cdots)$. If $j\le -N$,
then $2^{-j}x\in \bZ_2$,  which implies that $|2^{-j}x-a|_2>1$ for
all $a\in I_2$, $a\ne 0$.
Thus, $\phi\big(2^{-j}x-a\big)=0$ for all $a\in I_2$, $a\ne 0$,
and $\phi\big(2^{-j}x\big)=1$, whenever $j\le -N$.
Since, we have $f(x)=c_{j0}$ for all $j\le -N$.
Similarly, for another $x'=2^{-N'}(x_{0}'+x_{1}'2+x_{2}'2^2+\cdots)$,
we have $f(x')=c_{j'0}$ for all $j\le -N'$.
This yields that $f(x)=f(x')$. Consequently, $f(x)\equiv C$, where
$C$ is a constant. However, if $C\ne 0$, then $f\not\in \cL^2(\bQ_2)$.
Thus, $C=0$ what was to be proved.
\end{proof}

According to the above scheme, we introduce the space
$W_0$ as the orthogonal complement of $V_0$ in $V_{1}$.

Set
\begin{equation}
\label{73}
\psi^{(0)}(x)=\phi\Big(\frac{x}{2}\Big)
-\phi\Big(\frac{x}{2}-\frac{1}{2}\Big).
\end{equation}

\begin{Proposition}
\label{pr3.1}
The shift system  $\{\psi^{(0)}(\cdot-a), a\in I_2\}$, is
an orthonormal basis of the space $W_0$.
\end{Proposition}

\begin{proof}
Let us prove that $W_0\perp V_0$.
It follows from (\ref{62.0-4}), (\ref{73}) that
$$
\big(\psi^{(0)}(\cdot-a),\phi(\cdot-b)\big)
=\int_{\bQ_2}\psi^{(0)}(x-a)\phi(x-b)\,dx
\qquad\qquad\qquad\qquad\qquad
$$
$$
=\int_{\bQ_2}\bigg(\phi\Big(\frac{x}{2}-\frac{a}{2}\Big)
-\phi\Big(\frac{x}{2}-\frac{1}{2}-\frac{a}{2}\Big)\bigg)
\bigg(\phi\Big(\frac{x}{2}-\frac{b}{2}\Big)
+\phi\Big(\frac{x}{2}-\frac{1}{2}-\frac{b}{2}\Big)\bigg)\,dx
$$
for all $a,b\in I_2$. Let $a\ne b$. Since it is impossible
$a=b+1$ or $b=a+1$, taking into account that the functions
$2^{1/2}\phi(2^{-1}\cdot-c)$, $c\in I_2$ are mutually orthogonal,
we conclude that $\big(\psi^{(0)}(x-a),\phi(x-b)\big)=0$.
If $a=b$, again due to the orthonormality of the system
$\{2^{1/2}\phi(2^{-1}\cdot-c), c\in I_2\}$, taking into account
that $\frac{a}{2},\frac{a}{2}+\frac{1}{2}\in I_2$, we have
$$
\big(\psi^{(0)}(\cdot-a),\phi(\cdot-a)\big)
=\int_{\bQ_2}\bigg(\phi^2\Big(\frac{x}{2}-\frac{a}{2}\Big)
-\phi^2\Big(\frac{x}{2}-\frac{1}{2}-\frac{a}{2}\Big)\bigg)\,dx
\qquad\qquad
$$
$$
\qquad
=\int_{\bQ_2}\phi^2\Big(\frac{x}{2}-\frac{a}{2}\Big)\,dx
-\int_{\bQ_2}\phi^2\Big(\frac{x}{2}-\frac{1}{2}
-\frac{a}{2}\Big)\,dx=0.
$$
Thus, $\psi^{(0)}(\cdot+a)\perp\phi(\cdot+b)$ for  all $a,b\in I_2$.

Similarly, computing the integrals
$$
\big(\psi^{(0)}(\cdot-a), \psi^{(0)}(\cdot-b)\big)
=\int_{\bQ_2}\psi^{(0)}(x-a)\psi^{(0)}(x-b)\,dx
\qquad\qquad\qquad\qquad
$$
$$
=\int_{\bQ_2}\bigg(\phi\Big(\frac{x}{2}-\frac{a}{2}\Big)
-\phi\Big(\frac{x}{2}-\frac{1}{2}-\frac{a}{2}\Big)\bigg)
\bigg(\phi\Big(\frac{x}{2}-\frac{b}{2}\Big)
-\phi\Big(\frac{x}{2}-\frac{1}{2}-\frac{b}{2}\Big)\bigg)\,dx,
$$
we establish that the system $\{\psi^{(0)}(\cdot-a), a\in I_2\}$
is orthonormal.

It follows from
(\ref{62.0-4}) and (\ref{73})  that
$$
\phi\Big(\frac{x}{2}\Big)
=\frac{1}{2}\Big(\phi\big(x\big)+\psi^{(0)}\big(x\big)\Big),
\quad
\phi\Big(\frac{x}{2}-\frac{1}{2}\Big)
=\frac{1}{2}\Big(\phi\big(x\big)-\psi^{(0)}\big(x\big)\Big).
$$
If $a\in I_2$, then either $a=\frac12 b$, $b\in I_2$, or
$a=\frac12+\frac12 b$, $b\in I_2$. Hence,
$$
\phi\Big(\frac{x}{2}-a\Big)
=\frac{1}{2}\Big(\phi\big(x-b\big)+\psi^{(0)}\big(x-b\big)\Big),
\quad b\in I_2,
$$
whenever $a=\frac{1}{2}b$, and
$$
\phi\Big(\frac{x}{2}-a\Big)
=\frac{1}{2}\Big(\phi\big(x-b\big)-\psi^{(0)}\big(x-b\big)\Big),
\quad b\in I_2,
$$
whenever $a=\frac12+\frac12 b$.
Since $\{2^{1/2}\phi(2^{-1}\cdot-a):a\in I_2\}$ is a
basis for $V_1$, we obtain that the system
$\{\phi(\cdot-b), \psi^{(0)}(\cdot-b), b\in I_2\}$
is also a basis for $V_1$, i.e., the functions
$\psi^{(0)}(\cdot-b)$, $b\in I_2$, form a basis for the
space $W_0=V_{1}\ominus V_0$.
\end{proof}

Thus according to Propositions~\ref{pr2},~\ref{pr3},~\ref{pr3.1},
the collection $\{V_j:j\in\bZ\}$ is a MRA in
${\cL}^2(\bQ_2)$ and the function $\psi^{(0)}$ defined by (\ref{73})
is a wavelet function. This MRA is a $2$-adic analog of the real Haar
MRA and the wavelet basis generated by $\psi^{(0)}$ is an analog of
the real Haar basis. But in contrast to the real setting, the
{\it refinable function} $\phi$ generating our Haar MRA is
{\em periodic} with the period $1$ (see (\ref{62.0-0})), which
{\em never holds} for real refinable functions. It will be shown bellow that
due  this specific property of $\phi$, there exist infinity many different
orthonormal wavelet bases in the same Haar MRA (see Sec.~\ref{s5}).

Due to (\ref{8.2-1**}), (\ref{79}), the function $\psi^{(0)}$ can be
rewritten in the form
\begin{equation}
\label{62.0-7*}
\psi^{(0)}(x)=\chi_2(2^{-1}x)\Omega(|x|_2).
\end{equation}
Thus the Haar wavelet basis is
$$
\psi^{(0)}_{j a}(x)=2^{-j/2}\psi^{(0)}(2^{j}x-a)
\qquad\qquad\qquad\qquad\qquad\qquad\qquad\qquad\qquad
$$
\begin{equation}
\label{62.0-7}
=2^{-j/2}\chi_2\big(2^{-1}(2^{j}x-a)\big)
\Omega\big(|2^{j}x-a|_2\big), \quad x\in \bQ_2,
\quad j\in \bZ, \quad a\in I_2.
\end{equation}

Since a locally-constant function $\psi^{(0)}_{j a}(x)$ satisfies
the relation
\begin{equation}
\label{62.1-1}
\int_{\bQ_2}\psi^{(0)}_{j a}(x)\,dx=0,
\end{equation}
according to (\ref{54}), $\psi^{(0)}_{j a}(x)$ belongs to the
Lyzorkin space $\Phi(\bQ_2)$.

\begin{Remark}
\label{rem1} \rm
The Haar wavelet basis (\ref{62.0-7}) coincides with Kozyrev's
wavelet basis (\ref{62.0-1}) for the case $p=2$.
In present paper we restrict ourself by constructing the Haar wavelets
only for  $p=2$. Since Haar refinement equation (\ref{62.0-3}) was
presented for all $p$, a similar  construction may be easily  realized
in the general case.
Moreover, it is not difficult to see that Kozytev's wavelet function
$\theta_{j}(x)$ from (\ref{62.0-1}) can be expressed in terms of the
{\it refinable function} $\phi(x)$ as
\begin{equation}
\label{62.0}
\theta_{k}(x)=\chi_p(p^{-1}kx)\Omega\big(|x|_p\big)
=p^{-1/2}\sum_{r=0}^{p-1}h_r\phi\Big(\frac{1}{p}x-\frac{r}{p}\Big),
\quad x\in \bQ_p,
\end{equation}
where $h_r=p^{1/2}e^{2\pi i\{\frac{kr}{p}\}_p}$, $r=0,1,\dots,p-1$, \
$k=1,2,\dots,p-1$.
\end{Remark}

\begin{Remark}
\label{rem2} \rm
Because of periodicity (\ref{62.0-0}) of the refinable function
$\phi$, one can use the shifts  $\psi^{(0)}(\cdot+a)$, $a\in I_2$,
instead of  $\psi^{(0)}(\cdot-a)$, $a\in I_2$.
\end{Remark}

Now we show that there is another function $\psi^{(1)}$ whose shifts
form an orthonormal basis for $W_0$ (different from the basis generated
by $\psi^{(0)}$). Set
\begin{equation}
\label{73-1}
\psi^{(1)}(x)
=\frac{1}{\sqrt{2}}\bigg(\phi\Big(\frac{x}{2}\Big)
+\phi\Big(\frac{x}{2}-\frac{1}{2^2}\Big)
-\phi\Big(\frac{x}{2}-\frac{1}{2}\Big)
-\phi\Big(\frac{x}{2}-\frac{1}{2^2}-\frac{1}{2}\Big)
\bigg)
\end{equation}
and prove that the functions $\psi^{(1)}(\cdot-a)$, $a\in I_2$,
are mutually orthonormal. If  $a\in I_2$, $a\ne 0,\frac{1}{2}$,
then each of the numbers
$0,\frac{1}{2^2},\frac{1}{2},\frac{1}{2^2}+\frac{1}{2}$ differs
modulo $1$ from each of the numbers
$\frac{a}{2}, \frac{1}{2^2}+\frac{a}{2}, \frac{1}{2}+\frac{a}{2},
\frac{1}{2^2}+\frac{1}{2}+\frac{a}{2}$.
Due to orthonormality of the system  $\{2^{1/2}\phi(2^{-1}x-a), a\in I_2\}$
and (\ref{62.0-0}), it follows that $\psi^{(1)}$ is orthogonal
to $\psi^{(1)}(\cdot-a)$ whenever $a\in I_2$, $a\ne 0, \frac{1}{2}$.
Again due to orthonormality of the system
$\{2^{1/2}\phi(2^{-1}x-a), a\in I_2\}$
and (\ref{62.0-0}), we have
$$
\big(\psi^{(1)},\psi^{(1)}(\cdot-2^{-1})\big)
=\int_{\bQ_2}\psi^{(1)}(x)\psi^{(1)}(x-2^{-1})\,dx
\qquad\qquad\qquad\quad\qquad\qquad
$$
$$
=2^{-1}\int_{\bQ_2}\bigg(-\phi^2\Big(\frac{x}{2}\Big)
+\phi^2\Big(\frac{x}{2}-\frac{1}{2^2}\Big)
-\phi^2\Big(\frac{x}{2}-\frac{1}{2}\Big)
+\phi^2\Big(\frac{x}{2}-\frac{1}{2^2}-\frac{1}{2}\Big)
\bigg)\,dx=0,
$$
$$
\big(\psi^{(1)},\psi^{(1)}\big)
=\int_{\bQ_2}\psi^{(1)}(x)\psi^{(1)}(x)\,dx
\qquad\qquad\qquad\qquad\qquad\qquad\qquad\qquad\qquad
$$
$$
=2^{-1}\int_{\bQ_2}\bigg(\phi^2\Big(\frac{x}{2}\Big)
+\phi^2\Big(\frac{x}{2}-\frac{1}{2}\Big)
+\phi^2\Big(\frac{x}{2}-\frac{1}{2^2}-\frac{1}{2}\Big)
+\phi^2\Big(\frac{x}{2}-\frac{1}{2^2}\Big)\bigg)\,dx=1.
$$
Thus we proved that the system $\{\psi^{(1)}(\cdot+a), a\in I_2\}$ is
orthonormal.
It is not difficult to see that
\begin{equation}
\label{102-1}
\psi^{(1)}(x)=\frac{1}{\sqrt{2}}
\Big(\psi^{(0)}\big(x\big)+\psi^{(0)}\Big(x-\frac{1}{2}\Big)\Big),
\end{equation}
$$
\psi^{(1)}\Big(x-\frac{1}{2}\Big)=\frac{1}{\sqrt{2}}
\Big(-\psi^{(0)}\big(x\big)+\psi^{(0)}\Big(x-\frac{1}{2}\Big)\Big).
\qquad\quad
$$
This yields that
$$
\psi^{(0)}(x)=\frac{1}{\sqrt{2}}
\Big(\psi^{(1)}\big(x\big)-\psi^{(1)}\Big(x-\frac{1}{2}\Big)\Big).
$$
Since the system $\{\psi^{(0)}(\cdot-a), a\in I_2\}$ is
a  basis for  $W_0$, it follows that the system $\{\psi^{(1)}(\cdot-a),
a\in I_2\}$, is another orthonormal basis for $W_0$.

So, we showed that a wavelet basis generated by the Haar MRA
is {\em not unique}.

\section{Description of one-dimensional $2$-adic Haar bases}
\label{s5}

\subsection{Wavelet functions.}\label{s5.1}
Now we are going to show that there exist infinitely many
different wavelet functions $\psi^{(s)}$, $s\in \bN$, in
$W_0$ generating different bases for $\cL^2(\bQ_2)$.

In what foloows, we shall write the $2$-adic number
$a=2^{-s}\big(a_{0}+a_{1}2+\cdots+a_{s-1}2^{s-1}\big)\in I_2$, \
$a_{j}=0,1$, \ $j=0,1,\dots,s-1$, in the form $a=\frac{m}{2^s}$,
where $m=a_{0}+a_{1}2+\cdots+a_{s-1}2^{s-1}$.

Since the refinable function $\phi$ of the Haar MRA  is {\it
$1$-periodic} (see (\ref{62.0-0})), evidently, the wavelet
function $\psi^{0}$ has the following  property:
\begin{equation} \label{100} \psi^{(0)}(x\pm 1)=-\psi^{(0)}(x).
\end{equation}

Before we prove a general result, let us consider a simple special
case. Set
\begin{equation} \label{102}
\psi^{(1)}(x)=\alpha_{0}\psi^{(0)}(x)
+\alpha_{1}\psi^{(0)}\Big(x-\frac{1}{2}\Big), \quad
\alpha_{0},\alpha_{1}\in \bC,
\end{equation}
and find all the complex numbers $\alpha_{0}, \alpha_{1}$ for which
 $\{\psi^{(1)}(x-a), a\in I_2\}$ is
an orthonormal basis for  $W_0$.

Taking into account orthonormality of the system $\{\psi^{(0)}(\cdot-a),
a\in I_2\}$ and  (\ref{100}), we can easily see that $\psi^{(1)}$ is
orthogonal to $\psi^{(1)}(\cdot-a)$ whenever $a\in I_2$,
$a\ne 0, \frac{1}{2}$. Thus the system  $\{\psi^{(1)}(x-a), a\in I_2\}$
is orthonormal if and only if the system consisting  of the functions
(\ref{102}) and
\begin{equation}
\label{103}
\psi^{(1)}\Big(x-\frac{1}{2}\Big)=-\alpha_{1}\psi^{(0)}(x)
+\alpha_{0}\psi^{(0)}\Big(x-\frac{1}{2}\Big)
\end{equation}
is orthonormal, which is equivalent to the unitary property
of the matrix
$$
D=\left(
\begin{array}{cc}
\alpha_{0} & \alpha_{1} \\
-\alpha_{1} & \alpha_{0} \\
\end{array}
\right)
$$
It is clear that $D$ is a unitary matrix whenever
$|\alpha_{0}|^2+|\alpha_{1}|^2=1$. In this case
the system $\{\psi^{(1)}(\cdot-a),
a\in I_2\}$ is a basis for $W_0$ because $\{\psi^{(0)}(\cdot-a),
a\in I_2\}$ is a basis for $W_0$ and we have
$$
\psi^{(0)}(x)=\overline{\alpha_{0}}\psi^{(1)}(x)
-\overline{\alpha_{1}}\psi^{(1)}\Big(x-\frac{1}{2}\Big).
$$
So, $\psi^{(1)}$ is a Haar wavelet function if and only
if $|\alpha_{0}|^2+|\alpha_{1}|^2=1$. In particular, we
obtain (\ref{102-1}) for $\alpha_{0}=\alpha_{1}=\frac{1}{\sqrt{2}}$.

\begin{Theorem}
\label{th4}
Let $s=1,2,\dots$. The function
\begin{equation}
\label{101}
\psi^{(s)}(x)=\sum_{k=0}^{2^s-1}\alpha_{k}\psi^{(0)}\Big(x-\frac{k}{2^s}\Big),
\end{equation}
is a wavelet function for the Haar MRA if and only if
\begin{equation}
\label{108}
\alpha_{k}
=2^{-s}(-1)^{k}\sum_{r=0}^{2^s-1}\gamma_re^{-i\pi\frac{2r+1}{2^{s}}k},
\quad k=0,\dots,2^s-1, \quad\gamma_r\in \bC, \quad |\gamma_r|=1.
\end{equation}
\end{Theorem}

\begin{proof}
Let $\psi^{(s)}$ be defined by (\ref{101}).
Since  $\{\psi^{(0)}(\cdot-a), a\in I_2\}$ is an orthonormal system
(see Subsec.~\ref{s4.3}), taking into account (\ref{100}),
we see that $\psi^{(s)}$ is orthogonal to $\psi^{(s)}(\cdot-a)$
whenever $a\in I_2$, $a\ne \frac{k}{2^s}$, $k=0,1,\dots 2^s-1$. Thus
the system  $\{\psi^{(s)}(x-a), a\in I_2\}$ is orthonormal
if and only if the system  consisting of the functions
$$
\psi^{(s)}\Big(x-\frac{r}{2^s}\Big)=
-\alpha_{2^s-r}\psi^{(0)}(x)
-\alpha_{2^s-r+1}\psi^{(0)}\Big(x-\frac{1}{2^s}\Big)-\cdots
-\alpha_{2^s-1}\psi^{(0)}\Big(x-\frac{r-1}{2^s}\Big)
$$
\begin{equation}
\label{104}
\qquad
+\alpha_{0}\psi^{(0)}\Big(x-\frac{r}{2^s}\Big)+\cdots
+\alpha_{2^s-r-1}\psi^{(0)}\Big(x-\frac{2^s-1}{2^s}\Big),
\quad r=0,\dots,2^s-1,
\end{equation}
is orthonormal.
Set
$$
\Xi^{(0)}=\left(\psi^{(0)},
\psi^{(0)}\left(\cdot-\frac{1}{2^s}\right),\dots,
\psi^{(0)}\left(\cdot-\frac{2^s-1}{2^s}\right)\right)^T,
$$
$$
\Xi^{(s)}=\left(\psi^{(s)},
\psi^{(s)}\left(\cdot-\frac{1}{2^s}\right),\dots,
\psi^{(s)}\left(\cdot-\frac{2^s-1}{2^s}\right)\right)^T.
$$
By (\ref{104}), we have $\Xi^{(s)}=D\Xi^{(0)}$,
where
\begin{equation}
D=\label{105}
\left(
\begin{array}{cccccc}
\alpha_{0}&\alpha_{1}&\alpha_{2}&\ldots&\alpha_{2^s-2}&\alpha_{2^s-1} \\
-\alpha_{2^s-1}&\alpha_{0}&\alpha_{1}&\ldots&\alpha_{2^s-3}&\alpha_{2^s-2} \\
-\alpha_{2^s-2}&-\alpha_{2^s-1}&\alpha_{0}&\ldots&\alpha_{2^s-4}&
\alpha_{2^s-3} \\
\hdotsfor{6} \\
-\alpha_{2}&-\alpha_{3}&-\alpha_{4}&\ldots&\alpha_{0}&\alpha_{1} \\
-\alpha_{1}&-\alpha_{2}&-\alpha_{3}&\ldots&-\alpha_{2^s-1}&\alpha_{0} \\
\end{array}
\right).
\end{equation}
Due to orthonormality of  $\{\psi^{(0)}(\cdot-a), a\in I_2\}$,
the coordinates of  $\Xi^{(s)}$ form an orthonormal system
if and only if the matrix $D$ is unitary.

Let $u=(\alpha_{0},\alpha_{1},\dots,\alpha_{2^s-1})^{T}$ be a vector
and
$$
A=\left(
\begin{array}{cccccc}
0&0&\ldots&0&0&-1 \\
1&0&\ldots&0&0&0 \\
0&1&\ldots&0&0&0 \\
\hdotsfor{6} \\
0&0&\ldots&1&0&0 \\
0&0&\ldots&0&1&0 \\
\end{array}
\right).
$$
be a $2^s\times2^s$ matrix.
It is not difficult  to see that
$$
A^ru=\big(-\alpha_{2^s-r},-\alpha_{2^s-r+1},\dots,-\alpha_{2^s-1},
\alpha_{0},\alpha_{1},\dots,\alpha_{2^s-r-1}\big)^{T},
$$
where $r=1,2,\dots,2^s-1$. Thus $D=\big(u,Au,\dots,A^{2^s-1}u\big)^T$.
Hence, to describe all unitary matrixes $D$, we should find all
vectors $u=(\alpha_{0},\alpha_{1},\dots,\alpha_{2^s-1})^{T}$ such
that the system of vectors $\{A^ru, r=0,\dots,2^s-1\}$ is orthonormal.
We have already one such a vector $u_0=(1,0,\dots,0,0)^{T}$
because the matrix $D_0=\big(u_0,Au_0,\dots,A^{2^s-1}u_0\big)^T$
is the identity matrix. Let us prove that the system
$\{A^ru, r=0,\dots,2^s-1\}$ is orthonormal if and only if
$u=Bu_0$, where $B$ is a unitary matrix such that
$AB=BA$. Indeed, let $u=Bu_0$,  $B$ is a unitary matrix,
$AB=BA$. Then $A^ru=BA^ru_0$, \ $r=0,1,\dots,2^s-1$. Since the
system $\{A^ru_0, r=0,1,\dots,2^s-1\}$ is orthonormal and the
matrix $B$ is unitary, the vectors $A^ru$, \ $r=0,1,\dots,2^s-1$
are also orthonormal.
Conversely, if the system $A^ru$, \ $r=0,1,\dots,2^s-1$ is
orthonormal, taking into account that  $\{A^ru_0,
r=0,1,\dots,2^s-1\}$ is also an orthonormal system, we conclude that
there exists a unitary matrix $B$ such that $A^ru=B(A^ru_0)$, \
$r=0,1,\dots,2^s-1$. Since $A^{2^s}u=-u$, $A^{2^s}u_0=-u_0$,
we obtain  additionally  $A^{2^s}u=BA^{2^s}u_0$. It follows
from the above relations that $(AB-BA)(A^{r}u_0)=0$, \
$r=0,1,\dots,2^s-1$. Since the vectors $A^ru_0$, \
$r=0,1,\dots,2^s-1$ form a basis in the $2^s$-dimensional
space, we conclude that $AB=BA$.

Thus all  unitary matrixes $D$ are given by
$D=\big(Bu_0,BAu_0,\dots,BA^{2^s-1}u_0\big)^T$,
where $B$ is a unitary matrix such that $AB=BA$.
It remains to describe all such matrixes $B$.
It is not difficult to see that the eigenvalues of $A$ and
the corresponding normalized eigenvectors are  respectively
\begin{equation}
\label{106}
\lambda_r=-e^{i\pi\frac{2r+1}{2^{s}}}
\end{equation}
and $v_r=\big((v_r)_0,\dots,(v_r)_{2^s}\big)^T$,
where
\begin{equation}
\label{107}
(v_r)_l=2^{-s/2}(-1)^{l}e^{-i\pi\frac{2r+1}{2^{s}}l},
\quad l=0,1,2,\dots,2^s-1,
\end{equation}
$r=0,1,\dots,2^s-1$. Hence the matrix $A$ can be
represented as $A=C\widetilde{A}C^{-1}$, where
$$
\widetilde{A}=\left(
\begin{array}{ccccc}
\lambda_{0}&0&\ldots&0 \\
0&\lambda_{1}&\ldots&0 \\
\vdots&\vdots&\ddots&\vdots \\
0&0&\ldots&\lambda_{2^s-1} \\
\end{array}
\right)
$$
is a diagonal matrix, $C=\big(v_0,\dots,v_{2^s-1}\big)$
is a unitary matrix. It follows that the matrix $B=C\widetilde{B}C^{-1}$ is unitary if
and only if $\widetilde{B}$ is unitary. On the other hand, $AB=BA$ if
and only if $\widetilde{A}\widetilde{B}=\widetilde{B}\widetilde{A}$.
Moreover, since according to (\ref{106}), $\lambda_{k}\ne\lambda_{l}$
whenever $k\ne l$, all unitary matrix $\widetilde{B}$ such that
$\widetilde{A}\widetilde{B}=\widetilde{B}\widetilde{A}$, are given by
$$
\widetilde{B}=\left(
\begin{array}{ccccc}
\gamma_{0}&0&\ldots&0 \\
0&\gamma_{1}&\ldots&0 \\
\vdots&\vdots&\ddots&\vdots \\
0&0&\ldots&\gamma_{2^s-1} \\
\end{array}
\right),
$$
where $\gamma_k\in \bC$, $|\gamma_k|=1$. Hence all unitary matrix $B$
such that $AB=BA$, are given by $B=C\widetilde{B}C^{-1}$.
Using (\ref{107}), one can calculate
$$
\alpha_{k}=(Bu_0)_k=(C\widetilde{B}C^{-1}u_0)_k
=\sum_{r=0}^{2^s-1}\gamma_r(v_r)_k(\overline{v}_r)_0
\qquad\qquad\qquad\qquad\qquad
$$
$$
\qquad\quad
=2^{-s}(-1)^{k}\sum_{r=0}^{2^s-1}\gamma_re^{-i\pi\frac{2r+1}{2^{s}}k},
\quad k=0,1,\dots,2^s-1,
$$
where $\gamma_k\in \bC$, $|\gamma_k|=1$.

It remains to prove that the $\{\psi^{(s)}(\cdot-a),a\in I_2\}$ is a
basis for $W_0$ whenever $\psi^{(s)}$ is defined by (\ref{101}),
(\ref{108}). Since $\{\psi^{(0)}(\cdot-a),a\in I_2\}$ is a basis for
$W_0$, it suffices to check that any function $\psi^{(0)}(\cdot-c)$,
$c\in I_2$, can be decomposed with respect to the functions
$\psi^{(s)}(\cdot-a)$, $a\in I_2$. Any $c\in I_2$, $c\ne 0$, can be
represented in the form $c=\frac{r}{2^s}+b$, where $r=0,1,\dots,2^s-1$,
$|b|_2\ge 2^{s+1}$. Taking into account that $\Xi^{(0)}=D^{-1}\Xi^{(s)}$,
i.e.,
$$
\psi^{(0)}\Big(x-\frac{r}{2^s}\Big)
=\sum_{k=0}^{2^s-1}\beta_{k}^{(r)}\psi^{(s)}\Big(x-\frac{k}{2^s}\Big),
\quad r=0,1,\dots,2^s-1,
$$
we have
$$
\psi^{(0)}\Big(x-c\Big)=\psi^{(0)}\Big(x-\frac{r}{2^s}-b\Big)=
\sum_{k=0}^{2^s-1}\beta_{k}^{(r)}\psi^{(s)}\Big(x-\frac{k}{2^s}-b\Big),
$$

and $\frac{k}{2^s}+b\in I_2$, $k=0,1,\dots, 2^s-1$.
\end{proof}

Finally note  that, due to (\ref{54}),
$\int_{\bQ_2}\psi^{(s)}(2^jx-a)\,dx=0$, $j\in\bZ$, $a\in I_2$,
which yields that any function $\psi^{(s)}(2^j\cdot-a)$ belongs
to the Lizorkin space $\Phi(\bQ_2^n)$.

\subsection{Real wavelet functions.}\label{s5.2}
Using formulas (\ref{108}), one can extract all {\it real} wavelet
functions (\ref{101}).

Let $s=1$. According to (\ref{102}),  (\ref{103}),
\begin{equation}
\label{109}
\psi^{(1)}(x)=\cos\theta\,\psi^{(0)}\big(x\big)
+\sin\theta\,\psi^{(0)}\Big(x-\frac{1}{2}\Big)
\end{equation}
is the {\it real} wavelet function.

Let $s=2$. Set $\gamma_r=e^{i\theta_r}$, $r=0,1,\dots,2^s-1$.
It follows from (\ref{108})  that the wavelet function $\psi^{(1)}$
is real if and only if
$$
\begin{array}{rcl}
\displaystyle
\sin\theta_1+\sin\theta_2+\sin\theta_3+\sin\theta_4&=&0, \\
\displaystyle
\cos\theta_1-\cos\theta_2+\cos\theta_3-\cos\theta_4&=&0, \\
\displaystyle
\sin\theta_1-\sin\theta_2-\sin\theta_3+\sin\theta_4&=& \\
\displaystyle
\cos\theta_1+\cos\theta_2-\cos\theta_3-\cos\theta_4,&& \\
\displaystyle
\sin\theta_1-\sin\theta_2-\sin\theta_3+\sin\theta_4&=& \\
\displaystyle
-(\cos\theta_1+\cos\theta_2-\cos\theta_3-\cos\theta_4).&& \\
\end{array}
$$
These relations are equivalent to the system
$$
\begin{array}{rclrcl}
\displaystyle
\sin\theta_1&=&-\sin\theta_4, \quad \cos\theta_1&=&\cos\theta_4,\\
\displaystyle
\sin\theta_2&=&-\sin\theta_3, \quad \cos\theta_2&=&\cos\theta_3.\\
\end{array}
$$
Thus the real wavelet functions (\ref{101}) with $s=2$ are given by
$$
\psi^{(1)}(x)=\frac{1}{2}(\cos\theta_1+\cos\theta_2)\psi^{(0)}\big(x\big)
\qquad\qquad\qquad\qquad\qquad\qquad\qquad
$$
$$
+\frac{1}{2\sqrt{2}}(\cos\theta_1-\cos\theta_2+\sin\theta_1+\sin\theta_2)
\psi^{(0)}\Big(x-\frac{1}{2^2}\Big)
$$
$$
+\frac{1}{2}(\sin\theta_1-\sin\theta_2)\psi^{(0)}\Big(x-\frac{1}{2}\Big)
\qquad\qquad\qquad\qquad
$$
\begin{equation}
\label{110}
+\frac{1}{2\sqrt{2}}(\cos\theta_1-\cos\theta_2-\sin\theta_1-\sin\theta_2)
\psi^{(0)}\Big(x-\frac{1}{2^2}-\frac{1}{2}\Big).
\end{equation}
In particular, for the special cases $\theta_1=\theta_2=\theta$,
$\theta_1=-\theta_2=\theta$, $\theta_1=\theta_2+\frac{\pi}{2}=\theta$,
we obtain respectively the following one-parameter families of  real wavelet functions
\begin{equation}
\label{111}
\begin{array}{rclrcl}
\displaystyle
\psi^{(1)}(x)&=&\cos\theta\psi^{(0)}\big(x\big)
+\sin\theta\psi^{(0)}\Big(x-\frac{1}{2}\Big),\\
\displaystyle
\psi^{(1)}(x)&=&\cos\theta\psi^{(0)}\big(x\big)
+\frac{1}{\sqrt{2}}\sin\theta\psi^{(0)}\Big(x-\frac{1}{2^2}\Big) \\
\displaystyle
&& \qquad\qquad\qquad
-\frac{1}{\sqrt{2}}\sin\theta\psi^{(0)}\Big(x-\frac{1}{2^2}-\frac{1}{2}\Big),\\
\displaystyle
\psi^{(1)}(x)&=&\frac{1}{2}(\cos\theta-\sin\theta)\psi^{(0)}\big(x\big)
+\frac{1}{2\sqrt{2}}(\cos\theta+\sin\theta)\psi^{(0)}\Big(x-\frac{1}{2^2}\Big)\\
\displaystyle
&& \qquad\qquad\qquad\qquad\qquad
-\frac{1}{2}(\cos\theta-\sin\theta)\psi^{(0)}\Big(x-\frac{1}{2}\Big).\\
\end{array}
\end{equation}

\section{Description of multidimensional $2$-adic Haar bases}
\label{s6}

\subsection{$p$-Adic  separable multidimensional MRA}\label{s6.1}
Here we describe multidimensional wavelet bases constructed
by means of a tensor product of one-dimensional MRAs. This standard
approach for construction of multivariate  wavelets was suggested
by Y.~Meyer~\cite{13-1}  (see, e.g.,~\cite[\S 2.1]{NPS}).

Let $\{V_j^{(\nu)}\}_{j\in\bZ}$ , $\nu=1,\dots,n$, be one-dimensional
MRAs (see Subsec.~\ref{s4.1}).
We introduce subspaces $V_j$, $j\in\bZ$, of $\cL^2(\bQ_p^n)$ by
\begin{equation}
\label{d3}
V_j=\bigotimes_{\nu=1}^nV^{(\nu)}_j=
\overline{{\rm span}\{F=f_1\otimes\dots\otimes f_n, \ f_\nu\in V_j^{(\nu)}\}}.
\end{equation}
Let $\phi^{(\nu)}$ be a refinable function of $\nu$-th MRA $\{V_j^{(\nu)}\}_j$.
Set
\begin{equation}
\label{d3-1}
\Phi=\phi^{(1)}\otimes\dots\otimes\phi^{(n)}.
\end{equation}

Since the system $\{\phi^{(\nu)}(\cdot-a)\}_{a_{\nu}\in I_p}$ is an
orthonormal basis for $V^{(\nu)}_0$ (axiom~(e) of
Definition~\ref{de1}) for any $\nu=1,\dots,n$, it is clear that
$$
V_0=\overline{{\rm span}\{\Phi(\cdot-a): a=(a_1,\dots,a_n)\in I_p^n\}},
$$
where $I_p^n=I_p\times\cdots\times I_p$ is the direct product of $n$
sets $I_p$, and the system $\Phi(\cdot-a)$, $a\in I_p^n$, is an
orthonormal basis for $V_0$.
It follows from Definition (\ref{d3}) and  axiom~(d) of
Definition~\ref{de1} that $f\in V_0$ if and only if $f(2^{-j}\cdot)\in V_j$
for all $j\in\bZ$.
Since   axiom~(a) from Definition~\ref{de1} holds
for any one-dimensional MRA $\{V_j^{(\nu)}\}_j$, it is easy to see that
$\Phi(2^{-j}\cdot-a)\in V_{j+1}$ for any $a\in I_p^n$. Thus,
$V_j\subset V_{j+1}$.
It is not difficult to check that the axioms of completeness and
separability for the spaces $V_j$ hold. Thus we have the following statement.
\begin{Theorem}
\label{th1}
Let $\{V_j^{(\nu)}\}_{j\in\bZ}$, $\nu=1,\dots,n$, be KMAs in $\cL^2(\bQ_p)$.
Then the subspaces $V_j$ of  $\cL^2(\bQ_p^n)$ defined by~{\rm(\ref{d3})}
satisfy the following properties:

{\rm(a)} $V_j\subset V_{j+1}$ for all $j\in\bZ$;

{\rm(b)} $\cup_{j\in\bZ}V_j$ is dense in $\cL^2(\bQ_p^n)$;

{\rm(c)} $\cap_{j\in\bZ}V_j=\{0\}$;

{\rm(d)} $f(\cdot)\in V_j \Longleftrightarrow f(p^{-1}\cdot)\in V_{j+1}$
for all $j\in\bZ$;

{\rm(e)} the system $\{\Phi(x-a), a\in I_p^n\}$, is an orthonormal basis
for $V_0$, where $\Phi \in V_0$ is defined by {\rm(\ref{d3-1})}.
\end{Theorem}

Similarly to Definition~\ref{de1}, the collection of spaces
$V_j$, $j\in\bZ$, which satisfies  conditions (a)-(e) of Theorem~\ref{th1}
is called a {\it multiresolution analysis} in $\cL^2(\bQ_p^n)$,
the function $\Phi$ from axiom (e) is called refinable.

Next, following to the standard scheme (see, for example,~\cite[\S 2.1]{NPS}),
we define the wavelet spaces $W_j$ as the orthogonal complement of $V_j$ in
$V_{j+1}$, i.e.,
$$
W_j=V_{j+1}\ominus V_j, \quad j\in\bZ.
$$
Since
$$
V_{j+1}=\bigotimes\limits_{\nu=1}^nV_{j+1}^{(\nu)}=
\bigotimes\limits_{\nu=1}^n\big(V^{(\nu)}_j\oplus W^{(\nu)}_j\big)
\qquad\qquad\qquad\qquad
$$
$$
\qquad
=V_j\oplus
\bigoplus\limits_{e\subset\{1,\dots,n\}, \, e\ne\emptyset}
\big(\bigotimes\limits_{\nu\in e} W^{(\nu)}_j \big)
\big(\bigotimes\limits_{\mu\not\in e} V^{(\mu)}_j \big).
$$
So, the space $W_j$ is a direct sum of $2^n-1$ subspaces
$W_{j,e}$,  $e\subset\{1,\dots,n\}$, $e\ne\emptyset$.
Let $\psi^{(\nu)}$ be a wavelet function,
i.e. a function whose shifts (with respect to
$a\in I_p$) form an orthonormal basis for $W^{(\nu)}_0$. It is clear
that the shifts (with respect to
$a\in I_p^n$) of the function
\begin{equation}
\label{wd-62}
\Psi_e=\big(\bigotimes\limits_{\nu\in e} \psi^{(\nu)}\big)
\big(\bigotimes\limits_{\mu\not\in e}\phi^{(\mu)}\big),
\quad
e\subset\{1,\dots,n\}, \quad e\ne\emptyset,
\end{equation}
form an orthonormal basis for $W_{0,e}$.
So, we have
$$
\cL^2(\bQ_p^n)=\bigoplus\limits_{j\in\bZ}W_j
=\bigoplus\limits_{j\in\bZ}\Big(
\bigoplus\limits_{e\subset\{1,\dots,n\}, \, e\ne\emptyset}
W_{j,e}\Big),
$$
and  the  functions $p^{-nj/2}\Psi_e(p^j\cdot+a)$,
$e\subset\{1,\dots,n\}$, $e\ne\emptyset$, $j\in\bZ$, $a\in I_p^n$,
 form an orthonormal basis for $\cL^2(\bQ_p^n)$.

\subsection{Construction of multidimensional $2$-adic Haar MRA}\label{s6.2}
Let us apply the above construction taking the 2-adic Haar MRA
as $\nu$-th one-dimensional multiresolution analysis
$\{V_j^{(\nu)}\}_{j\in\bZ}$, $\nu=1,\dots,n$.

To construct multivariate wavelet functions (\ref{wd-62}),
we choose $\psi^{(0)}$ as a wavelet function for each
one-dimensional MRA. Thus we have the following $2^n-1$
multidimensional wavelet functions
$$
\begin{array}{lll}
\displaystyle
\Psi^{(0)}_{\{1,\dots,n\}}=\psi^{(0)}(x_1)\psi^{(0)}(x_2)\cdots
\psi^{(0)}(x_{n-1})\psi^{(0)}(x_n),&& \\
\displaystyle
\Psi^{(0)}_{\{1,\dots,n-1\}}=\psi^{(0)}(x_1)\psi^{(0)}(x_2)\cdots
\psi^{(0)}(x_{n-1})\phi(x_n), &&\\
\hdotsfor{1} \\
\displaystyle
\Psi^{(0)}_{\{2,\dots,n\}}=\phi(x_1)\psi^{(0)}(x_2)\cdots
\psi^{(0)}(x_{n-1})\psi^{(0)}(x_n), && \\
\hdotsfor{1} \\
\hdotsfor{1} \\
\hdotsfor{1} \\
\displaystyle
\Psi^{(0)}_{\{1\}}=\psi^{(0)}(x_1)\phi(x_2)\cdots\phi(x_{n-1})\phi(x_n), && \\
\hdotsfor{1} \\
\displaystyle
\Psi^{(0)}_{\{n\}}=\phi(x_1)\phi(x_2)\cdots\phi(x_{n-1})\psi^{(0)}(x_n). && \\
\end{array}
$$

Let $e\subset\{1,\dots,n\}$, $e\ne\emptyset$. Denote by
$k_{e}=\big((k_{e})_1,\dots,(k_{e})_n\big)$ the vector
whose coordinates are given by
$$
(k_{e})_{\nu}=\left\{
\begin{array}{lcr}
1, &&\quad \nu \in e, \\
0, && \nu \not\in e, \\
\end{array}
\right.
\ \ \ \nu=1,\dots,n.
$$
Since $\phi(x_{\nu})=\Omega\big(|x_{\nu}|_2\big)$ and
$\psi^{(0)}(x_{\nu})=\chi_2(2^{-1}x_{\nu})\Omega(|x_{\nu}|_2)$, \
$x_{\nu}\in \bQ_p$, $\nu=1,2,\dots,n$, (see (\ref{62.0-4}) and
(\ref{62.0-7*})), it follows from (\ref{9}), (\ref{62.0-4}),
that the wavelet function $\Psi^{(0)}_{e}$ can be rewritten as
\begin{equation}
\label{w-62.8}
\Psi^{(0)}_{e}(x)=\chi_2\big(2^{-1}k_{e}\cdot x\big)
\Omega\big(|x|_2\big), \quad x=(x_1,\dots,x_n)\in \bQ_2^n.
\end{equation}

According to the above consideration, we have we  the following statement
\begin{Theorem}
\label{th4.0}
The system of functions
$$
\Psi^{(0)}_{e;j a}(x)=2^{-nj/2}\Psi^{(0)}_{e}(2^{j}x-a)
\qquad\qquad\qquad\qquad\qquad\qquad\qquad\qquad\quad
$$
\begin{equation}
\label{w-62.8=1}
=2^{-nj/2}\chi_2\big(2^{-1}k_{e}\cdot (2^{j}x-a)\big)
\Omega\big(|2^{j}x-a|_2\big), \quad x\in \bQ_2^n,
\end{equation}
$e\subset\{1,\dots,n\}$, $e\ne\emptyset$, \ $j\in \bZ$, $a\in I_2^n$,
is an orthonormal  basis for ${\cL}^2(\bQ_2^n)$.
\end{Theorem}

Now we construct multidimensional wavelet bases using
different one- dimensional Haar wavelet bases (see Subsec.~\ref{s5.1}).
Namely, we apply the construction of Subsec.~\ref{s6.1}
taking again the Haar MRA as $\nu$-th one-dimensional multiresolution
analysis $\{V_j^{(\nu)}\}_{j\in\bZ}$, $\nu=1,\dots,n$, and choosing
wavelet functions $\psi^{(s_{\nu})}$ for construction of  multivariate wavelet functions
(\ref{wd-62}).

Let $s=(s_1,\dots,s_n)$, where $s_{\nu}\in \bN_0$, $\nu=1,2,\dots,n$.
We have the following $2^n-1$ wavelet functions
$$
\begin{array}{lll}
\displaystyle
\Psi^{(s)}_{\{1,\dots,n\}}(x)=\psi^{(s_1)}(x_1)\psi^{(s_2)}(x_2)\cdots
\psi^{(s_{n-1})}(x_{n-1})\psi^{(s_n)}(x_n), && \\
\displaystyle
\Psi^{(s)}_{\{1,\dots,n-1\}}(x)=\psi^{(s_1)}(x_1)\psi^{(s_2)}(x_2)\cdots
\psi^{(s_{n-1})}(x_{n-1})\phi(x_n), && \\
\hdotsfor{1} \\
\displaystyle
\Psi^{(s)}_{\{2,\dots,n\}}(x)=\phi(x_1)\psi^{(s_2)}(x_2)\cdots
\psi^{(s_{n-1})}(x_{n-1})\psi^{(s_n)}(x_n), && \\
\hdotsfor{1} \\
\hdotsfor{1} \\
\hdotsfor{1} \\
\displaystyle
\Psi^{(s)}_{\{1\}}(x)=\psi^{(s_1)}(x_1)\phi(x_2)\cdots\phi(x_{n-1})\phi(x_n), && \\
\hdotsfor{1} \\
\displaystyle
\Psi^{(s)}_{\{n\}}(x)=\phi(x_1)\phi(x_2)\cdots\phi(x_{n-1})\psi^{(s_n)}(x_n).
\end{array}
$$

Set $\alpha_{r}^{1}=\alpha_{r}$, where $\alpha_{r}$
is given by (\ref{108}), $r=0,1,\dots,2^{s}-1$, and $\alpha_{0}^{0}=1$,
$\alpha_{1}^{0}=\cdots\alpha_{2^{s}-1}^{0}=0$.
Since $\phi(x_{\nu})=\Omega\big(|x_{\nu}|_2\big)$, $x_{\nu}\in \bQ_2$,
and $\psi^{(s_{\nu})}$ is given by (\ref{101}), (\ref{108}),
$\nu=1,2,\dots,n$, due to (\ref{9}), the wavelet functions
$\Psi^{(s)}_{e}$ can be rewritten as
$$
\Psi^{(s)}_{e}(x)=
\qquad\qquad\qquad\qquad\qquad\qquad\qquad\qquad\qquad
\qquad\qquad\qquad\qquad\qquad\quad
$$
\begin{equation}
\label{62.8}
=\sum_{r_1=0}^{2^{s_1}-1}\cdots\sum_{r_n=0}^{2^{s_n}-1}
\alpha_{r_1}^{(k_{e})_1}\cdots\alpha_{r_n}^{(k_{e})_n}
\Psi^{(0)}_{e}
\Big(x-\Big(\frac{r_1}{2^{s_1}}(k_{e})_1,\dots,\frac{r_n}{2^{s_n}}(k_{e})_n\Big)\Big),
\quad x \in \bQ_2^n,
\end{equation}
where $\Psi^{(0)}_{e}$ is defined by (\ref{w-62.8}),
$e\subset\{1,\dots,n\}$, $e\ne\emptyset$.

According to the above consideration, we have the following statement.
\begin{Theorem}
\label{th4.1}
The system of functions
$$
\Psi^{(s)}_{e;j a}(x)=
\qquad\qquad\qquad\qquad\qquad\qquad\qquad\qquad\qquad
\qquad\qquad\qquad\qquad\quad
$$
\begin{equation}
\label{62.9}
=\sum_{r_1=0}^{2^{s_1}-1}\cdots\sum_{r_n=0}^{2^{s_n}-1}
\alpha_{r_1}^{(k_e)_1}\cdots\alpha_{r_n}^{(k_e)_n}
\Psi^{(0)}_{e}
\Big(2^{j}x-a
-\Big(\frac{r_1}{2^{s_1}}(k_e)_1,\dots,\frac{r_n}{2^{s_n}}(k_e)_n\Big)\Big),
\end{equation}
$x \in \bQ_2^n$, \ $e\subset\{1,\dots,n\}$, $e\ne\emptyset$, \
$j\in\bZ$, $a\in I_2^n$ forms an orthonormal  basis for
${\cL}^2(\bQ_2^n)$.
\end{Theorem}

Since for a locally-constant function $\Psi^{(s)}_{e;j a}(x)$
the relation $\int_{\bQ_2^n}\Psi^{(s)}_{e;j a}(x)\,d^nx=0$ holds,
in view of (\ref{54}), $\Psi^{(s)}_{e;j a}\in \Phi(\bQ_2^n)$.

\section{$p$-Adic wavelets as eigenfunctions of pseudo-differential operators}
\label{s8}

\subsection{Pseudo-differential operators.}\label{s8.1}
S.~V.~Kozyrev~\cite{Koz0} proved that one-dimensional $p$-adic
wavelets (\ref{62.0-1}) are eigenfunctions of the Vladimirov
fractional operator $D^{\alpha}$, $\alpha>0$. In fact,  this
statement holds for all $\alpha\in \bC$ due to (\ref{62.1-1}).
 A criterion
for pseudo-differential operators (\ref{64.3}) to have wavelets
(\ref{62.0-1}) and (\ref{62.0-1*}) as eigenfunctions
was found in~\cite{Al-Kh-Sh3},~\cite{Kh-Sh1}.

Now we consider a similar problem for $2$-adic wavelets (\ref{w-62.8=1})
and (\ref{62.9}).

\begin{Theorem}
\label{th5}
Let $A$ be a pseudo-differential operator {\rm (\ref{64.3})}
with a symbol $\cA\in \cE(\bQ_2^n\setminus \{0\})$,
\ $e\subset\{1,\dots,n\}$, $e\ne\emptyset$, \ $j\in \bZ$, $a\in I_2^n$.
Then the  function
$$
\Psi^{(0)}_{e;j a}(x)=2^{-nj/2}\chi_2\big(2^{-1}k_e\cdot (2^{j}x-a)\big)
\Omega\big(|2^{j}x-a|_2\big), \quad x\in \bQ_2^n,
$$
is an eigenfunction of $A$ if and only if
\begin{equation}
\label{64.1***}
\cA\big(2^{j}(-2^{-1}k_e+\eta)\big)=\cA\big(-2^{j-1}k_e\big),
\qquad \forall \, \eta \in \bZ_2^n.
\end{equation}
The corresponding eigenvalue is $\lambda=\cA\big(-2^{j-1}k_e\big)$, i.e.,
$$
A\Psi^{(0)}_{e;j a}=\cA(-2^{j-1}k_e)\Psi^{(0)}_{e;j a}.
$$
\end{Theorem}

\begin{proof}
Combining  (\ref{14}),
(\ref{14.1}), (\ref{9}) with (\ref{w-62.8=1}),
we obtain
$$
F[\Psi^{(0)}_{e;j a}](\xi)=2^{-nj/2}F[\Psi^{(0)}_{e}(2^{j}x-a)](\xi)
=2^{nj/2}\chi_2\big(2^{-j}a\cdot\xi\big)
F[\Psi^{(0)}_{e}(x)](2^{-j}\xi)
$$
$$
\qquad
=2^{nj/2}\chi_2\big(2^{-j}a\cdot\xi\big)
F\Big[\prod_{\nu=1}^n\chi_2(2^{-1}(k_e)_{\nu}x_{\nu})
\Omega(|x_{\nu}|_2)\Big](2^{-j}\xi)
$$
$$
\quad
=2^{nj/2}\chi_2\big(2^{-j}a\cdot\xi\big)
\prod_{\nu=1}^nF\Big[\Omega(|x_\nu|_2)\Big]
(2^{-1}(k_e)_{\nu}+2^{-j}\xi_{\nu})
$$
\begin{equation}
\label{64.8*}
=2^{nj/2}\chi_2\big(2^{-j}a\cdot\xi\big)
\Omega\big(|2^{-1}k_e+2^{-j}\xi|_2\big).
\qquad\qquad\qquad
\end{equation}
It is clear that $\Omega\big(|2^{-1}(k_e)_{\nu}+\xi_{\nu}|_2\big)\ne 0$
only if $\xi_k=-2^{-1}(k_e)_{\nu}+\eta_{\nu}$, where $\eta_{\nu}\in \bZ_2$, \
$\nu=1,2,\dots,n$. This yields $\xi=-2^{-1}k_e+\eta$,
$\eta \in \bZ_p^n$.

If condition (\ref{64.1***}) is satisfied, then, using (\ref{64.3}),
(\ref{64.8*}), we have
$$
A\Psi^{(0)}_{e;j a}(x)=F^{-1}\big[\cA(\xi)F[\Psi^{(0)}_{e;j a}(\xi)\big](x)
\qquad\qquad\qquad\qquad\qquad\qquad\qquad\quad
$$
\begin{equation}
\label{64.9}
=2^{nj/2}F^{-1}\big[\cA(\xi)\chi_2\big(2^{-j}a\cdot\xi\big)
\Omega\big(|2^{-1}k_e+2^{-j}\xi|_2\big)\big](x).
\end{equation}
Making the change of variable $\xi\to 2^{j}(-2^{-1}k_e+\eta)$
and using (\ref{14.1}), we obtain
$$
A\Psi^{(0)}_{e;j a}(x)
=2^{-nj/2}
\int\limits_{\bQ_2^n}\chi_2\big(-(2^{j}x-a)\cdot(-2^{-1}k_e+\eta)\big)
\cA(2^{j}(-2^{-1}k_e+\eta))\,\Omega(|\eta|_2)\,d^n\eta
$$
$$
\qquad\quad
=2^{-nj/2}\cA(-2^{j-1}j)
\chi_2\big(2^{-1}k_e\cdot(2^{j}x-a)\big)
\int_{B_{0}^n}\chi_2(-(2^{j}x-a)\cdot\eta)\,d^n\eta
$$
$$
=\cA(-2^{j-1}j)\Psi^{(0)}_{e;j a}(x).
\qquad\qquad\qquad\qquad\qquad\qquad\qquad\qquad
$$
Consequently, $A\Psi^{(0)}_{e;j a}(x)=\lambda\Psi^{(0)}_{e;j a}(x)$,
where $\lambda=\cA(-2^{j-1}k_e)$.

Conversely, if $A\Psi^{(0)}_{e;j a}=\lambda\Psi^{(0)}_{e;j a}$,
$\lambda\in \bC$, taking the Fourier transform from
both the left and the right hand sides of this identity and
using (\ref{64.8*}), (\ref{64.9}), we have
\begin{equation}
\big(\cA(\xi)-\lambda\big)\chi_2\big(2^{-j}a\cdot\xi\big)
\Omega\big(|2^{-1}k_e+2^{-j}\xi|_2\big)=0, \quad \xi\in \bQ_2^n.
\label{9.5}
\end{equation}
If now $\eta \in \bZ_2^n$, $\xi=2^{j}(-2^{-1}k_e+\eta)$, then
$2^{-1}k_e+2^{-j}\xi=\eta\in \bZ_2^n$. Since
$\Omega\big(|2^{-1}k_e+2^{-j}\xi|_2\big)\ne0$ and
$\chi_2\big(2^{-j}a\cdot\xi\big)\ne0$, it follows from
(\ref{9.5}) that $\lambda=\cA(\xi)$.
Thus $\lambda=\cA\big(2^{j}(-2^{-1}k_e+\eta)\big)$ for any
$\eta \in \bZ_2^n$. In particular, $\lambda=\cA(-2^{j-1}j)$,
and, consequently, (\ref{64.1***}) holds.
\end{proof}

In particular, we have the following statement.

\begin{Corollary}
\label{cor5}
Let $e\subset\{1,\dots,n\}$, $e\ne\emptyset$, \ $j\in \bZ$, $a\in I_2^n$.
Then the function $\Psi^{(0)}_{e;j a}$ is an eigenfunction of the
fractional operator {\rm (\ref{59**})}.
The corresponding eigenvalue is $\lambda=2^{\alpha(1-j)}$, i.e.,
$$
D^{\alpha}\Psi^{(0)}_{e;j a}=2^{\alpha(1-j)}\Psi^{(0)}_{e;j a},
\quad \alpha \in\bC.
$$
\end{Corollary}

\begin{proof}
The symbol $\cA(\xi)=|\xi|_2^{\alpha}$ of the fractional operator
$D^\alpha$ satisfies condition (\ref{64.1***}):
$$
\cA\big(2^{j}(-2^{-1}k_e+\eta)\big)
=\big|2^{j}(-2^{-1}k_e+\eta)\big|_2^{\alpha}
=2^{-j\alpha}
\big(\max_{1\le \nu\le n}\big|-2^{-1}(k_e)_{\nu}+\eta_{\nu}\big|_2\big)^{\alpha}
$$
$$
\qquad\qquad\qquad
=2^{\alpha(1-j)}
\big(\max_{1\le \nu\le n}\big|2^{j}(-2^{-1}(k_e)_{\nu})\big|_2\big)^{\alpha}
=\cA(-2^{j-1}k_e)
$$
for all $\eta \in \bZ_2^n$, \ $e\subset\{1,\dots,n\}$, $e\ne\emptyset$.
Here we take into account that $(k_e)_{\nu}=0,1$; $\nu=1,2,\dots,n$;
$(k_e)_1+\cdots+(k_e)_n\ne 0$.
Thus, by Theorem~\ref{th5}, $\Psi^{(0)}_{e;j a}$ is an eigenfunction and
the corresponding eigenvalue is $\lambda=2^{\alpha(1-j)}$.
\end{proof}

Similarly to Theorem~\ref{th5} and Corollary~\ref{cor5},
using representation (\ref{62.9}),
it is not difficult to prove the following statements.
\begin{Theorem}
\label{th6}
Let $A$ be a pseudo-differential operator {\rm (\ref{64.3})} with
a symbol $\cA(\xi)\in \cE(\bQ_p^n\setminus \{0\})$,
  $s=(s_1,\dots,s_n)$, where $s_{\nu}\in \bN_0$,
 $e\subset\{1,\dots,n\}$, $e\ne\emptyset$, \ $j\in \bZ$, $a\in I_2^n$. Then the
 function $\Psi^{(s)}_{e;j a}$
 is an eigenfunction of $A$ if and
only if {\rm (\ref{64.1***})} holds.
The corresponding eigenvalue is $\lambda=\cA\big(-2^{j-1}k_e\big)$, i.e.,
$$
A\Psi^{(s)}_{e;j a}=\cA(-2^{j-1}k_e)\Psi^{(s)}_{e;j a}.
$$
\end{Theorem}

\begin{Corollary}
\label{cor6}
Let  $s=(s_1,\dots,s_n)$, where $s_{\nu}\in \bN_0$,
$e\subset\{1,\dots,n\}$, $e\ne\emptyset$, \ $j\in \bZ$, $a\in I_2^n$.
Then the function $\Psi^{(s)}_{e;j a}$ is an eigenfunction of
the fractional operator {\rm (\ref{59**})}.
The corresponding eigenvalue is $\lambda=2^{\alpha(1-j)}$, i.e.,
$$
D^{\alpha}\Psi^{(s)}_{e;j a}(x)
=2^{\alpha(1-j)}\Psi^{(s)}_{e;j a}(x),
\quad \alpha \in\bC.
$$
\end{Corollary}

\section*{Acknowledgments}

The authors are greatly indebted to E.~Yu.~Panov for fruitful discussions.

\end{document}